\documentclass[12pt]{article}
\usepackage{setspace}
\usepackage{amsthm}
\usepackage{amsmath,amsfonts,amssymb}
\usepackage{bm}
\usepackage{natbib}
\usepackage{changepage}
\usepackage{graphicx}
\usepackage{caption}
\usepackage{subcaption}
\usepackage{rotating}
\usepackage{pdflscape}
\usepackage[utf8]{inputenc}
\usepackage[english]{babel}

\usepackage{authblk}
\usepackage{paralist,algorithmic,algorithm}

\usepackage{booktabs}
\usepackage{multirow}
\usepackage{makecell}
\usepackage{color}
\usepackage{placeins}
\usepackage{float}
\usepackage{mathabx}
\usepackage{afterpage}
\usepackage[export]{adjustbox}
\usepackage{epstopdf}
\usepackage{url}
\usepackage[pdftex, plainpages=false, pdfpagelabels]{hyperref}
\hypersetup{
    linktocpage=true,
    colorlinks=true,
    citecolor=blue,
    urlcolor=blue,
    linkcolor=blue,
    citebordercolor={1 0 0},
    breaklinks=true,
    }

\usepackage{tikz}
\usetikzlibrary{shapes.geometric, arrows, positioning, quotes, calc, decorations.pathreplacing, calligraphy, fit}
\tikzstyle{cohort} = [rectangle, minimum height=18pt, minimum width=48pt, align=center, draw=black, fill=white!10]
\tikzstyle{arrow} = [thick,->,>=stealth]

\usepackage{array} 
\newcolumntype{L}[1]{>{\raggedright\arraybackslash}p{#1}}
\newcolumntype{R}[1]{>{\raggedleft\arraybackslash}p{#1}}

\catcode`\^ = 13 \def^#1{\sp{#1}{}}

\definecolor{Pink}{rgb}{1.0, 0.5, 0.5}
\definecolor{Maroon}{rgb}{0.8, 0.0, 0.0}

\def\boxit#1{\vbox{\hrule\hbox{\vrule\kern6pt\vbox{\kern6pt#1\kern6pt}\kern6pt\vrule}\hrule}}

\newtheorem{theorem}{Theorem}[section]
\newtheorem{lemma}[theorem]{Lemma}
\newtheorem{proposition}[theorem]{Proposition}

\newtheorem{assumption}{Assumption}

\newcommand{\bM}{\mbox{\bf M}}

\newcommand{\bb}{\mbox{\bf b}}

\newcommand{\be}{\mbox{\bf e}}

\newcommand{\bu}{\mbox{\bf u}}

\newcommand{\bw}{\mbox{\bf w}}
\newcommand{\bx}{\mbox{\bf x}}
\newcommand{\by}{\mbox{\bf y}}

\newcommand{\bA}{\mbox{\bf A}}
\newcommand{\bB}{\mbox{\bf B}}
\newcommand{\bC}{\mbox{\bf C}}

\newcommand{\bD}{\mbox{\bf D}}
\newcommand{\bE}{\mbox{\bf E}}

\newcommand{\bH}{\mbox{\bf H}}

\newcommand{\bI}{\mbox{\bf I}}

\newcommand{\bL}{\mbox{\bf L}}
\newcommand{\bl}{\mbox{\bf l}}
\newcommand{\bQ}{\mbox{\bf Q}}

\newcommand{\bU}{\mbox{\bf U}}
\newcommand{\bV}{\mbox{\bf V}}
\newcommand{\bW}{\mbox{\bf W}}
\newcommand{\bX}{\mbox{\bf X}}

\newcommand{\bY}{\mbox{\bf Y}}
\newcommand{\bZ}{\mbox{\bf Z}}

\newcommand{\bbeta}{\mbox{\boldmath $\beta$}}
\newcommand{\btheta}{\mbox{\boldmath $\theta$}}
\newcommand{\bTheta}{\mbox{\boldmath $\Theta$}}

\newcommand{\bSigma}{\mbox{\boldmath $\Sigma$}}

\newcommand{\bmu}{\mbox{\boldmath $\mu$}}
\newcommand{\bphi}{\mbox{\boldmath $\phi$}}

\def\t{^\top}

\def\beqn{\begin{eqnarray}}
\def\eeqn{\end{eqnarray}}
\def\beqns{\begin{eqnarray*}}
\def\eeqns{\end{eqnarray*}}

\def\0{{\bf 0}}
\def\A{{\bf A}}

\def\C{{\bf C}}

\def\t{{\bf t}}

\def\bP{{\bf P}}
\def\bQ{{\bf Q}}

\def\X{{\bf X}}

\def\Y{{\bf Y}}

\def\1{{\bf 1}}

\newcommand{\plogis}{\mathrm{plogis}}

\newcommand{\blind}{0}

\def\trans{^{\rm T}}
\def\strans{^{*\rm T}}

\def\t1trans{^{t+1\rm T}}

\def\spacingset#1{\renewcommand{\baselinestretch}%
{#1}\small\normalsize}
\spacingset{1}

\begin{document}

\title{\bf Salvaging Forbidden Treasure in Medical Data: Utilizing Surrogate Outcomes and Single Records for Rare Event Modeling}


 \if1\blind
 {
   \author{}\date{}
   \maketitle
 } \fi

 \if0\blind
 {
  \author{Xiaohui Yin$^1$, Shane Sacco$^{2}$, Robert H. Aseltine$^{2,3}$, Fei Wang$^4$, Kun Chen$^{1,2}$\thanks{Corresponding author; kun.chen@uconn.edu}\\    
$^1$\textit{Department of Statistics, University of Connecticut (UConn)}\\
$^2$\textit{Center for Population Health, UConn Health Center (UCHC)}\\
$^3$\textit{Division of Behavioral Sciences and Community Health, UCHC}\\
$^4$\textit{Department of Population Health Sciences, Weill Cornell Medical College}
}
 \maketitle
 }
 \fi




\begin{abstract}

\singlespacing
The vast repositories of Electronic Health Records (EHR) and medical claims hold untapped potential for studying rare but critical events, such as suicide attempt. Conventional setups often model suicide attempt as a univariate outcome and also exclude any ``single-record'' patients with a single documented encounter due to a lack of historical information. However, patients who were diagnosed with suicide attempts at the only encounter could, to some surprise, represent a substantial proportion of all attempt cases in the data, as high as 70--80\%. We innovate a hybrid \& integrative learning framework to leverage concurrent outcomes as surrogates and harness the ``forbidden" yet precious information from single-record data. Our approach employs a supervised learning component to learn the latent variables that connect primary (e.g., suicide) and surrogate outcomes (e.g., mental disorders) to historical information. It simultaneously employs an unsupervised learning component to utilize the single-record data, through the shared latent variables. As such, our approach offers a general strategy for information integration that is crucial to modeling rare conditions and events.
With hospital inpatient data from Connecticut, we demonstrate that single-record data and concurrent diagnoses indeed carry valuable information, and utilizing them can substantially improve suicide risk modeling.\\ 

\end{abstract}

\doublespacing

\section{Introduction}\label{sec:intro}

Electronic Health Records (EHR) data and medical claims data represent transformative resources in modern healthcare, offering a comprehensive repository of patient information that spans demographics, medical diagnoses, laboratory measures, and clinical outcomes. The significance of such real-world medical data lies in its rich, longitudinal nature, providing a detailed narrative of a patient's healthcare journey. Utilizing this wealth of information is pivotal for advancing medical research and improving patient care. 

Researchers exhibit a growing interest in understanding and predicting rare conditions through the lens of large-scale medical data. One prominent example is the prediction of suicide or suicidal behavior; these are rare but devastating events with profound public health implications. Statistical modeling of such events from EHR or medical claims data can facilitate early intervention strategies and potentially save lives. Several studies have demonstrated promise in identifying individuals at alleviated risk for suicide, thereby enabling timely and targeted preventive measures \citep{Barak-Corren2017, Simon2018}. In recent years, the growing aggregation of medical data has provided a fertile ground for developing sophisticated models capable of discerning subtle patterns indicative of elevated suicide risk \citep{Walsh2018, Kessler2019, DoshiChen2020, Su2020TransPsy}. 

However, despite recent strides in the modeling of suicide risk using large-scale medical data, the estimation and prediction performance of these rare event models seems to have reached a plateau \citep{Belsher2019}. The efficacy of these models can vary significantly across different clinical settings, data availability, and data utilization \citep{Torok2019, ZangHou2023}. Persistent challenges remain in this domain, even as it opens up new avenues for innovation. For example, several studies considered utilizing external datasets to capture critical information such as social determinants of health \citep{chen2020social, xu2022improving, lybarger2023leveraging}. 

In this study, we take a ``data-centric" approach and aim to improve rare event modeling by better utilizing existing data at hand. First of all, to alleviate the rarity of suicide outcomes, we consider utilizing other co-current and related outcomes, such as suicide ideation, mental health problems and other related comorbidities; these surrogates could potentially enrich the information on the targeted rare outcome. Another particular challenge or opportunity encountered in suicide risk studies is the prevalence of ``single-record'' data \citep{Kessler2015}. These are from individuals who have only a single encounter recorded in the available data, which include those whose only recorded encounter was due to a suicide attempt without any prior medical records. Surprisingly, this group can represent a substantial proportion of patients with records of suicide attempt \citep{Tran2019}. For example, in our motivating study with the Connecticut Hospital Inpatient Discharge Database (HIDD), among 40,702 pediatric patients, we identified 1,895 suicide attempt cases, of which 1,408 cases or over 74\% were from single-record without any historical information! 
In the EHR data from the Kansas Health Information Network (KHIN) for patients between 18 and 64 years of age and the period between January 1, 2014, and December 31, 2015, there were 216 and 1,218 patients identified to have had suicide attempts with and without historical information, respectively, which means almost 85\% of the attempt cases were from single-record data. Such problems are also commonly encountered when modeling other health outcomes with EHR data. Unfortunately, in statistical analysis and machine learning, these single-record patients are often excluded from the model-building process due to the absence of historical medical records to inform parameter estimation and prediction \citep{Riblet2017}. This exclusion appears inevitable, yet it raises concerns about the comprehensiveness and representativeness of the resulting models, as they may fail to account for and utilize a significant subset of the at-risk population.

We propose a hybrid and integrative learning framework to enrich the suicide outcomes and harness the often ``forbidden'' yet precious information from single-record data. Addressing these gaps could potentially enhance estimation and prediction accuracy and clinical utility of risk models \citep{Franklin2017}.

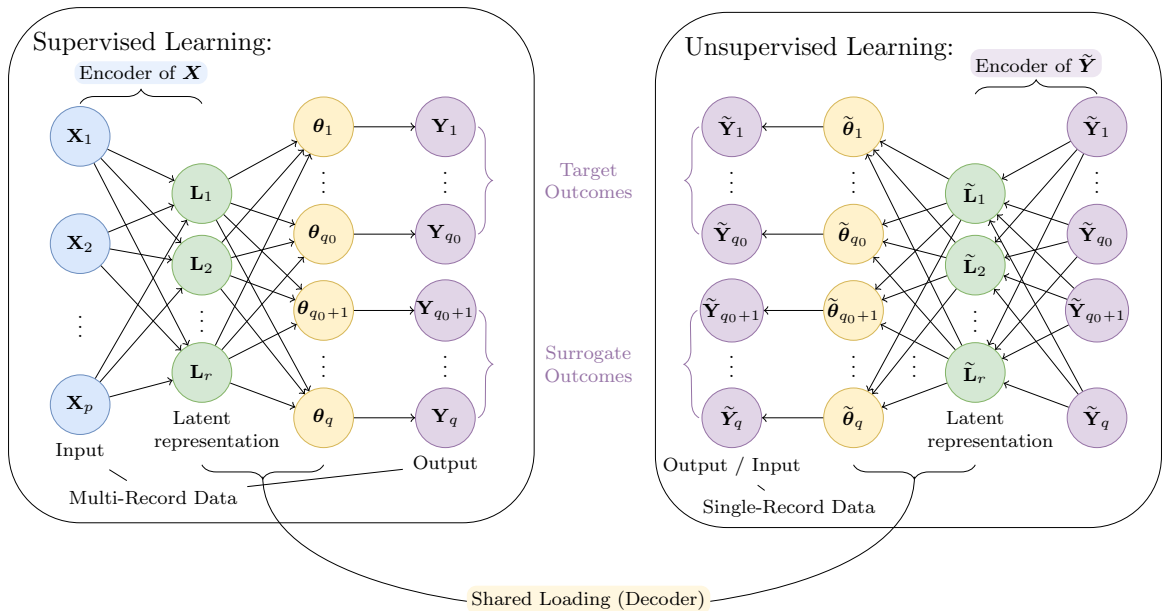
\begin{figure}[htp]
\begin{center}
  \resizebox{0.95\textwidth}{!}{  
    \begin{tikzpicture}[node distance=32pt, decoration={calligraphic brace, mirror, amplitude=6pt}, font=\scriptsize]
    \pgfmathsetmacro{\nodedist}{30pt} 
    \pgfmathsetmacro{\columnnodedist}{1.7*\nodedist}
    \pgfmathsetmacro{\onehalfnodedist}{1.5*\nodedist}
    \pgfmathsetmacro{\doublenodedist}{2*\nodedist}
    
    \definecolor{darkpurple}{HTML}{9673A6}
    \definecolor{lightpurple}{HTML}{E1D5E7}
    \definecolor{darkblue}{HTML}{6C8EBF}
    \definecolor{lightblue}{HTML}{DAE8FC}
    \definecolor{darkgreen}{HTML}{82B366}
    \definecolor{lightgreen}{HTML}{D5E8D4}
    \definecolor{darkyellow}{HTML}{D6B656}
    \definecolor{lightyellow}{HTML}{FFF2CC}
    \tikzstyle{var_X} = [circle, minimum height=25pt, align=center, draw=black, draw=darkblue, fill=lightblue, text=black]
    \tikzstyle{var_Y} = [circle, minimum height=25pt, align=center, draw=black, draw=darkpurple, fill=lightpurple, text=black, inner sep=0pt]
    \tikzstyle{var_L} = [circle, minimum height=25pt, align=center, draw=black, draw=darkgreen, fill=lightgreen, text=black]
    \tikzstyle{var_Theta} = [circle, minimum height=25pt, align=center, draw=black, draw=darkyellow, fill=lightyellow, text=black, inner sep=0pt]
    \tikzstyle{arrow} = [thick,->,>=stealth]

    \node (X1) [var_X] {$\bX_1$};
    \node (X2) [var_X, below of=X1, node distance = \onehalfnodedist] {$\bX_2$};
    \node (Xp) [var_X, below of=X2, node distance = 1.5 * \onehalfnodedist] {$\bX_p$};
    
    \node (L1) [var_L, right of=X1, node distance = \columnnodedist, yshift = -24pt] {$\bL_1$};
    \node (L2) [var_L, below of=L1, node distance = \nodedist] {$\bL_2$};
    \node (Lr) [var_L, below of=L2, node distance = \onehalfnodedist] {$\bL_r$};
    
    \node (theta1) [var_Theta, right of=L1, node distance = \columnnodedist, yshift = 28pt] {$\btheta_1$};
    \node (thetaq0) [var_Theta, below of=theta1, node distance = \onehalfnodedist] {$\btheta_{q_0}$};
    \node (thetaq0plus1) [var_Theta, below of=thetaq0] {$\btheta_{q_0+1}$};
    \node (thetaq) [var_Theta, below of=thetaq0plus1, node distance = \onehalfnodedist] {$\btheta_q$};
    
    \node (Y1) [var_Y, right of=theta1, node distance = \columnnodedist] {$\bY_1$};
    \node (Yq0) [var_Y, below of=Y1, node distance = \onehalfnodedist] {$\bY_{q_0}$};
    \node (Yq0plus1) [var_Y, below of=Yq0] {$\bY_{q_0+1}$};
    \node (Yq) [var_Y, below of=Yq0plus1, node distance = \onehalfnodedist] {$\bY_q$};
    
    \node (tY1) [var_Y, right of=Y1, node distance = 2 * \doublenodedist] {$\widetilde{\bY}_1$};
    \node (tYq0) [var_Y, below of=tY1, node distance = \onehalfnodedist] {$\widetilde{\bY}_{q_0}$};
    \node (tYq0plus1) [var_Y, below of=tYq0] {$\widetilde{\bY}_{q_0+1}$};
    \node (tYq) [var_Y, below of=tYq0plus1, node distance = \onehalfnodedist] {$\widetilde{\bm Y}_q$};
    
    \node (ttheta1) [var_Theta, right of=tY1, node distance = \columnnodedist] {$\widetilde{\btheta}_1$};
    \node (tthetaq0) [var_Theta, below of=ttheta1, node distance = \onehalfnodedist] {$\widetilde{\btheta}_{q_0}$};
    \node (tthetaq0plus1) [var_Theta, below of=tthetaq0] {$\widetilde{\btheta}_{q_0+1}$};
    \node (tthetaq) [var_Theta, below of=tthetaq0plus1, node distance = \onehalfnodedist] {$\widetilde{\btheta}_q$};
    
    \node (tL1) [var_L, right of=ttheta1, node distance = \columnnodedist,  yshift = -28pt] {$\widetilde{\bL}_1$};
    \node (tL2) [var_L, below of=tL1, node distance = \nodedist] {$\widetilde{\bL}_2$};
    \node (tLr) [var_L, below of=tL2, node distance = \onehalfnodedist] {$\widetilde{\bL}_r$};
    
    \node (tY1_prime) [var_Y, right of=tL1, node distance = \columnnodedist, yshift = 28pt] {$\widetilde{\bY}_1$};
    \node (tYq0_prime) [var_Y, below of=tY1_prime, node distance = \onehalfnodedist] {$\widetilde{\bY}_{q_0}$};
    \node (tYq0plus1_prime) [var_Y, below of=tYq0_prime] {$\widetilde{\bY}_{q_0+1}$};
    \node (tYq_prime) [var_Y, below of=tYq0plus1_prime, node distance = \onehalfnodedist] {$\widetilde{\bY}_q$};

    \foreach \nnname in {theta,Y,ttheta,tY} {
        \node at ($(\nnname 1)!0.5!(\nnname q0)$)[yshift=2pt] {$\vdots$};
        \node at ($(\nnname q0plus1)!0.5!(\nnname q)$)[yshift=2pt] {$\vdots$};
    }
    \node at ($(tY1_prime)!0.5!(tYq0_prime)$)[yshift=2pt] {$\vdots$};
    \node at ($(tYq0plus1_prime)!0.5!(tYq)$)[yshift=2pt] {$\vdots$};
    \node at ($(X2)!0.5!(Xp)$)[yshift=2pt] {$\vdots$};
    \node at ($(L2)!0.5!(Lr)$)[yshift=2pt] {$\vdots$};
    \node at ($(tL2)!0.5!(tLr)$)[yshift=2pt] {$\vdots$};

    \node (InputSupervised) at (Xp) [below=12pt] {{Input}};
    \node at (Lr) [below=12pt] {\parbox{40pt}{\centering Latent\\ representation}};
    \node (OutputSupervised) at (Yq) [below=12pt] {{Output}};
    
    \node (OutputUnsupervised) at (tYq) [below=12pt] {{Output / Input}};
    \node at (tLr) [below=12pt] {\parbox{40pt}{\centering Latent\\ representation}};
    
    \coordinate (middle) at ($(InputSupervised)!0.2!(OutputSupervised)$); 
    \node (MultiData) [below of = middle, node distance = 0.6 * \nodedist] {Multi-Record Data};
    \node (SingleData) [below of = OutputUnsupervised, xshift = 24pt, node distance = 0.6 * \nodedist] {Single-Record Data};
    \draw[-] (MultiData) -- (InputSupervised);
    \draw[-] (MultiData) -- (OutputSupervised);
    \draw[-] (SingleData) -- (OutputUnsupervised);
    
    \node (Supervised) [above of = X1, node distance = \onehalfnodedist, font = \small, anchor=west, yshift=-6pt, xshift=-24pt] {Supervised Learning:};
    \node (Unsupervised) [above of = tY1, node distance = \onehalfnodedist, font = \small, anchor=west, yshift=-12pt, xshift=-24pt] {Unsupervised Learning:};

    \draw[decorate, decoration ={brace,raise=1pt}] ([xshift=- \columnnodedist, yshift=-6pt]thetaq.south) -- ([yshift=-6pt]thetaq.south) node [midway,xshift=-0.8cm] {};
    \draw[decorate, decoration ={brace,raise=1pt}]([yshift=-6pt]tthetaq.south) -- ([xshift=\columnnodedist, yshift=-6pt]tthetaq.south) node [midway,xshift=-0.8cm] {};
    \draw[-, looseness = 0.7] ([yshift=-12pt, xshift=-0.5 * \columnnodedist]thetaq.south) to[out=270, in=270] 
    node[midway, above, fill=lightyellow!60, draw=none, rounded corners, inner sep=2pt, yshift=-2pt] {Shared Loading (Decoder)} ([yshift=-12pt, xshift=0.5 * \columnnodedist]tthetaq.south);

    \draw[decorate, decoration={brace, raise=-1pt}] ([xshift=\columnnodedist, yshift=2pt]X1.north) -- ([yshift=2pt]X1.north) node[midway, yshift=12pt, fill=lightblue!60, draw=none, rounded corners, inner sep=2pt] {Encoder of $\bm X$};
    \draw[decorate, decoration ={brace,raise=-1pt}]([yshift=2pt]tY1_prime.north) -- ([xshift = -\columnnodedist, yshift=2pt]tY1_prime.north) node [midway, yshift=12pt, fill=lightpurple!60, draw=none, rounded corners, inner sep=2pt] {Encoder of $\widetilde{\bm Y}$};

    \draw[decorate, decoration={brace, raise=-1pt}, color=darkpurple] ([xshift=2pt]Yq0.east) --([xshift=2pt]Y1.east) node(target1) [midway,xshift=12pt]{};
    \draw[decorate, decoration={brace, raise=-1pt}, color=darkpurple] ([xshift=-2pt]tY1.west)--([xshift=-2pt]tYq0.west) node(target2) [midway,xshift=-12pt]{};

    \draw[decorate, decoration={brace, raise=-1pt}, color=darkpurple] ([xshift=2pt]Yq.east) -- ([xshift=2pt]Yq0plus1.east) node(surrogate1) [midway,xshift=12pt]{};
    \draw[decorate, decoration={brace, raise=-1pt}, color=darkpurple] ([xshift=-2pt]tYq0plus1.west)--([xshift=-2pt]tYq.west) node(surrogate2) [midway,xshift=-12pt]{};

    \node[color=darkpurple, text width=42pt, align=center] at ($(target1)!0.5!(target2)$) {Target Outcomes};
    \node[color=darkpurple, text width=42pt, align=center] at ($(surrogate1)!0.5!(surrogate2)$) {Surrogate Outcomes};

    \draw [rounded corners=30pt] ([yshift = 6pt, xshift = -6pt]Supervised.north west) -- ([yshift = -7*\nodedist, xshift = -6pt]Supervised.north west)--([yshift = -7*\nodedist, xshift = 4.2*\columnnodedist]Supervised.north west) -- ([yshift = 6pt, xshift = 4.2*\columnnodedist]Supervised.north west) -- cycle;
    \draw [rounded corners=30pt] ([yshift = 6pt, xshift = -8pt]Unsupervised.north west) -- ([yshift = -7*\nodedist, xshift = -8pt]Unsupervised.north west)--([yshift = -7*\nodedist, xshift = 4*\columnnodedist]Unsupervised.north west) -- ([yshift = 6pt, xshift = 4*\columnnodedist]Unsupervised.north west) -- cycle;

    \foreach \i in {1,2,p} {
        \foreach \k in {1,2,r}{
                \draw[->] (X\i) -- (L\k);
        }
    }
    
    \foreach \k in {1,2,r} {
        \foreach \j in {1,q0,q0plus1,q}{
                \draw[->] (L\k) -- (theta\j);
                \draw[->] (tL\k) -- (ttheta\j);
                \draw[->] (tY\j_prime) -- (tL\k);
        }
    }
    
    \foreach \j in {1,q0,q0plus1,q} {
        \draw[->] (theta\j) -- (Y\j);
        \draw[->] (ttheta\j) -- (tY\j);
    }

  \end{tikzpicture}
}
\end{center}
    \caption{Schematic representation of the hybrid \& integrative learning framework. }
        \label{fig:diag_semi_rr}
\end{figure}

Our methodology unfolds in two key ideas: (1) integrated learning with both primary and surrogate outcomes and (2) hybrid learning with both multi-record and single-record data. Figure \ref{fig:diag_semi_rr} depicts a schematic representation of the proposed framework. First and foremost, it is recognized that some co-occurring diagnoses with the primary outcome of suicide, particularly those on mental health, can, to some extent, act as surrogate outcomes to suicide. An integrative model that treats both suicide and these co-occurring conditions as multivariate outcomes could facilitate information borrowing from these interrelated outcomes. Specifically, we posit that suicide and other co-occurring conditions are linked to historical records through a small number of latent factors, leading to the adoption of a latent-variable model or latent-representation learning as the supervised learning component of our framework. 
Second, with the integrated outcomes, the problem becomes multivariate and enables the incorporation of single-record data within this framework. We recognize that suicide and co-occurring conditions could maintain a low-dimensional structure even in the absence of historical information. Thus, a low-rank encoding of these outcomes can serve as the unsupervised learning component, in which the latent variables are self-learned without the supervision of historical records and yet are expected to resemble those learned from the supervised component. The final framework is thus a hybrid of supervised and unsupervised learning components. We remark that this setup should be distinguished from semi-supervised learning \citep{van_engelen_survey_2020}, where the unsupervised learning is typically on the unlabeled features rather than the integrated outcomes.

Our proposed hybrid approach aims to capitalize on the intricate relationships between suicide and co-current medical conditions, employing a combination of supervised and unsupervised learning techniques to bolster the estimation and predictive accuracy of suicide risk based on historical medical records. It's important to note that the framework we've introduced is not limited to suicide study alone; its versatility allows for broad applicability in modeling various rare conditions. Furthermore, the framework can cope with either classical or more advanced methodologies, such as reduced-rank regression \citep{anderson1956, izenman1975, reinsel2022} or neural networks. Specifically, the supervised and unsupervised components could be parameterized as two generalized linear low-rank models with shared latent factors, which could be seamlessly substituted with two neural networks, each sharing an internal layer of hidden nodes. This shared layer would embody the latent variables, ensuring that the interconnectedness of the outcomes is preserved and leveraged. We demonstrate that our proposed model led to substantially improved estimation and prediction in the suicide risk study with HIDD data. 

The rest of the paper is organized as follows. Section~\ref{sec:data} provides a descriptive analysis of the HIDD data for studying suicide risk, which motivated us to consider utilizing surrogate outcomes and single-record data. Section~\ref{sec:model} presents our hybrid \& integrative learning framework, under a general encoder-decoder modeling setup. Computational algorithms and some theoretical considerations are presented in Section~\ref{sec:computation}, and simulation studies are reported in Section~\ref{sec:simulation}. In Section~\ref{sec:application}, we apply the proposed approach to predict suicide attempts with the HIDD data from the State of Connecticut as a proof of concept. Some concluding remarks and future directions are provided in  Section~\ref{sec:discussion}.


\section{Data Description \& Exploratory Analysis}\label{sec:data}

We focus our study on a cohort of pediatric, adolescent, and young adult inpatients ranging from 10 to 24 years of age, drawn from the Connecticut Hospital Inpatient Discharge Database (HIDD). HIDD provides a comprehensive repository of inpatient hospitalization records from all acute care hospitals in Connecticut, offering a rich medical data source. Our goal is to predict the first occurrence of a suicide attempt in patients based on available data from October 1, 2012 to September 30, 2017. Suicide attempts were identified by ICD-9 (International Classification of Diseases, Version 9) codes designating intentional self-harm and combinations of codes strongly indicative of suicide attempts \citep{haerian2012methods,xu2022improving}. 

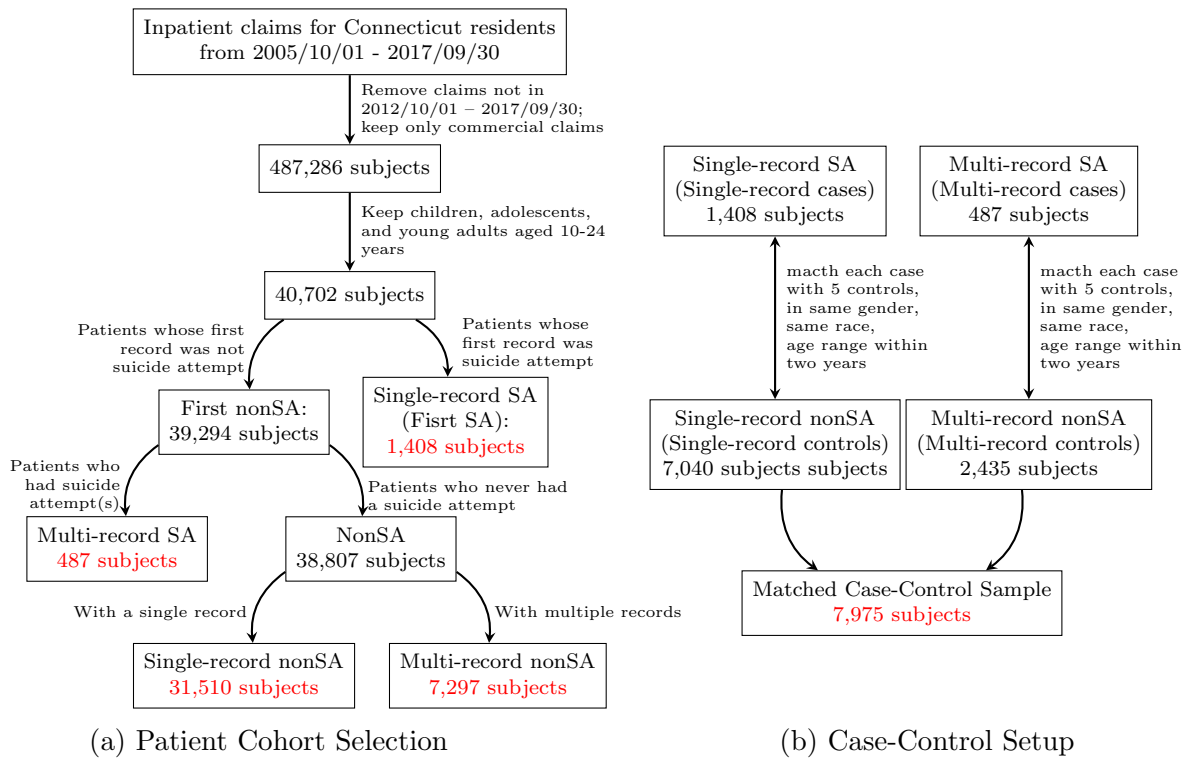
\begin{figure}[htp]
\centering
\begin{subfigure}{0.45\textwidth}
\begin{tikzpicture}[node distance=48pt]
\node (start) [cohort, font=\scriptsize] { Inpatient claims for Connecticut residents\\from 2005/10/01 - 2017/09/30};
\node (screen) [cohort, font=\scriptsize, below of=start] {487,286 subjects};
\node (children) [cohort, font=\scriptsize, below of=screen] {40,702 subjects};
\node (first_non_sa) [cohort, font=\scriptsize, below of=children, xshift=-40pt] {First nonSA:\\39,294 subjects};
\node (single_record_sa) [cohort, font=\scriptsize, below of=children, xshift=40pt] {Single-record SA\\(Fisrt SA):\\\color{red}{1,408 subjects}};
\node (multi_record_sa) [cohort, font=\scriptsize, below of=first_non_sa, xshift=-48pt] {Multi-record SA \\\color{red}{487 subjects}};
\node (non_sa) [cohort, font=\scriptsize, below of=first_non_sa, xshift=48pt] {NonSA\\38,807 subjects};
\node (single_record_non_sa) [cohort, font=\scriptsize, below of=non_sa, xshift=-48pt] {Single-record nonSA\\\color{red}{31,510 subjects}};
\node (multi_record_non_sa) [cohort, font=\scriptsize, below of=non_sa, xshift=48pt] {Multi-record nonSA\\\color{red}{7,297 subjects}};


\draw [arrow] (start) -- node[right] {\tiny \parbox{100pt}{\raggedright Remove claims not in 2012/10/01 – 2017/09/30;\\keep only commercial claims}}(screen);

\draw [arrow] (screen) -- node[right] {\tiny \parbox{100pt}{\raggedright Keep children, adolescents, and young adults aged 10-24 years}} (children);

\draw [arrow, bend right=30] (children) to node[above, left] {\tiny \parbox{70pt}{\raggedleft Patients whose first record was not suicide attempt }} (first_non_sa);
\draw [arrow, bend left=30] (children) to node[above, right, xshift=4pt] {\tiny \parbox{60pt}{\raggedright Patients whose first record was suicide attempt}} (single_record_sa);

\draw [arrow, bend right=30] (first_non_sa) to node[above, left, yshift = -4pt] {\tiny \parbox{45pt}{\raggedleft Patients who had suicide attempt(s)}}  (multi_record_sa);

\draw [arrow, bend left=30] (first_non_sa) to node[above, right, yshift = -8pt] {\tiny \parbox{76pt}{\raggedright Patients who never had a suicide attempt}} (non_sa);

\draw [arrow, bend right=30] (non_sa) to node[above, left, yshift = -4pt] {\tiny \parbox{80pt}{\raggedleft With a single record}} (single_record_non_sa);

\draw [arrow, bend left=30] (non_sa) to node[above, right, yshift = -4pt] {\tiny \parbox{80pt}{\raggedright With multiple records}} (multi_record_non_sa);

\end{tikzpicture}
\caption{Patient Cohort Selection}
\end{subfigure}
~
\hspace{0.05\textwidth}
\begin{subfigure}{0.45\textwidth}
\begin{tikzpicture}
\node (single_record_sa) [cohort, font=\scriptsize] {Single-record SA\\(Single-record cases)\\1,408 subjects};
\node (multi_record_sa) [cohort, font=\scriptsize, right of=single_record_sa, node distance=96pt] {Multi-record SA\\(Multi-record cases)\\487 subjects};

\node (single_record_non_sa) [cohort, font=\scriptsize,
below of=single_record_sa, node distance=96pt] {Single-record nonSA\\(Single-record controls)\\7,040 subjects subjects};
\node (multi_record_non_sa) [cohort, font=\scriptsize, below of=multi_record_sa, node distance=96pt] {Multi-record nonSA\\(Multi-record controls)\\2,435 subjects};

\coordinate (middle) at ($(single_record_non_sa)!0.5!(multi_record_non_sa)$); 

\node (matched) [cohort, font=\scriptsize, below of=middle, node distance=60pt] {Matched Case-Control Sample\\\color{red}{7,975 subjects}};

\node (blank) [below of=matched,node distance=36pt] {};


\draw [thick,<->,>=stealth] (single_record_sa) to node[above, right] {\tiny \parbox{64pt}{\raggedright macth each case\\
with 5 controls,\\in same gender,\\same race,\\age range within two years}} (single_record_non_sa);

\draw [thick,<->,>=stealth] (multi_record_sa) to node[above, right] {\tiny \parbox{64pt}{\raggedright macth each case\\
with 5 controls,\\in same gender,\\same race,\\age range within two years}} (multi_record_non_sa);

\draw [arrow, bend left=30] (multi_record_non_sa) to (matched);

\draw [arrow, bend right=30] (single_record_non_sa) to (matched);

\end{tikzpicture}
\caption{Case-Control Setup}
\end{subfigure}
\caption{Cohort selection and case-control setup with HIDD data.}\label{fig:hidd_cohort}
\end{figure}

Figure~\ref{fig:hidd_cohort} describes the cohort selection and case-control setup with the HIDD data. From Figure~\ref{fig:hidd_cohort}(a), the cohort consisted of patients with inpatient commercial claims between October 1, 2005, and September 30, 2017, at 10 to 24 years of age. A substantial number of patients with a single record were identified: within the cohort of 40,702 patients, a noteworthy total of 32,918 presented with only one record, in which 1,408 individuals were identified with a suicide attempt. In contrast, among 7784 patients with at least one historical record, only 487 patients were identified with a suicide attempt. As such, a majority of suicide attempt cases would be discarded if conventional modeling approaches were used \citep{Barak-Corren2017,tsui2021natural,xu2022improving}. In Figure~\ref{fig:hidd_cohort}(b), we matched the cases and controls on gender, race, and age range within two years, with a 1:5 case-control ratio; the final case-control cohort consisted of 7,975 patients.

It is important to explore the main characteristics of the patient groups with different data configurations. The 7,975 patients can be categorized into four groups based on the number of records: single-record cases (those with one record that was a suicide attempt), single-record controls (those with one record that was not a suicide attempt), multi-record cases (those with a suicide attempt record and with at least one historical record), and multi-record controls (those with multiple records and without any suicide attempt record). A demographic breakdown is presented in Table~\ref{tab:hidd_cohort_demo}. Interestingly, the group of the multi-record cases and that of the single-record cases show similar distributions in gender, race, and age, e.g., the percentage of females is 63.66\%, and 67.61\%, the percentage of white is 78.64\% and 73.22\%, and the percentage for 15-19 age group is 50.72\% and 50.50\%, for multiple-record and single-record cases, respectively. All three differences are not significant at $0.01$ significance level based on $\chi^2$ tests. This alleviates the concern that the single-record cases fundamentally differ from the multi-record counterparts.

\begin{table}[H]
\singlespacing
\centering 
\scriptsize
\caption{Descriptive characteristics of the case-control data. The percentages are not shown for sex and race in the controls, as they are matched exactly with those of the cases. 
}
\begin{tabular}[t]{l r r r r}
\toprule
 & Multi-record cases & Multi-record controls & Single-record cases & Single-record controls \\
\midrule

Total & 487 & 2435 & 1408 & 7040  \\

Sex &  &  &  & \\
\hspace{2em}Female & 310 (63.66\%) & 1505 & 952 (67.61\%) & 4760 \\
\hspace{2em}Male & 177 (36.34\%) & 885 & 456 (32.39\%) & 2280 \\

Race &  &  &  & \\
\hspace{2em}Asian & 12 (2.46\%) & 60 & 27 (1.92\%) & 135 \\
\hspace{2em}Black & 36 (7.39\%) & 180 & 106 (7.53\%) & 530 \\
\hspace{2em}Hisp & 32 (6.57\%) & 160 & 116 (8.24\%) & 580 \\
\hspace{2em}Other & 24 (4.93\%) & 120 & 128 (9.09\%) & 640\\
\hspace{2em}White & 383 (78.64\%) & 1915  & 1031 (73.22\%) & 5155 \\

Age &  &  &  & \\
\hspace{2em} 10-14 & 46 (9.45\%) & 237 (9.73\%) & 176 (12.5\%) & 830 (11.79\%)\\
\hspace{2em} 15-19 & 247 (50.72\%) & 1161 (47.68\%) & 711 (50.5\%) & 3449 (48.99\%)\\
\hspace{2em} 20-24 & 194 (39.84\%) & 1037 (42.59\%) & 521 (37\%) & 2761 (39.22\%)\\
\bottomrule
\end{tabular}
\label{tab:hidd_cohort_demo}
\end{table}

To realize our integrative learning framework in Figure~\ref{fig:diag_semi_rr}, we need to select a set of surrogate outcomes to be analyzed jointly with the primary outcome of interest. In the suicide risk study, our consideration of choosing the surrogates gravitates towards concurrent mental disorders, which, by their nature, have strong connections with incidents of suicide attempts \citep{mitra2023associations}. Table~\ref{tab:surrogate_summary} reports the prevalence of these conditions in the cases and controls. Previous studies indicated that these disorders are significantly associated with suicide attempts, after adjusting for demographics and social determinants \citep{blosnich2020social}; the details of their identification are given in Appendix~\ref{sec:apd:mental_health}. The case patients are much more likely to experience mental health disorders than the controls; for example, among multi-record cases, 58.73\% and 36.34\% patients experienced major depressive disorder and anxiety disorder, respectively, while among multi-record controls, the corresponding proportions reduced to 25.79\% and 18.15\%. On the other hand, comparing the multi-record cases and the single-record cases, the prevalence of mental disorders is quite similar in general. 

The above exploratory analysis underpins our strategy of pursuing a latent low-dimensional structure to represent suicide attempts and concurrent mental health conditions jointly. Since this structure can be shared and learned from either multi-record or single-record data, unleashing the ``forbidden'' power within the single-record data becomes possible.

\begin{table}[H]
\singlespacing
\centering
\scriptsize
\caption{Summary of concurrent mental health disorders in the case-control data.}
\label{tab:surrogate_summary}
\begin{tabular}[t]{l r r r r}
\toprule
Outcomes & Multi-record cases & Multi-record controls & Single-record cases & Single-record controls\\
\midrule
Major Depressive Disorder & 286 (58.73\%) & 628 (25.79\%) & 791 (56.18\%) & 1386 (19.69\%) \\

Alcohol Use Disorder & 50 (10.27\%) & 114 (4.68\%) & 187 (13.28\%) & 317 (4.5\%) \\

Drug Use Disorder & 126 (25.87\%) & 348 (14.29\%) & 300 (21.31\%) & 668 (9.49\%) \\

Anxiety Disorder & 177 (36.34\%) & 442 (18.15\%) & 432 (30.68\%) & 885 (12.57\%) \\

Posttraumatic Stress Disorder & 66 (13.55\%) & 178 (7.31\%) & 128 (9.09\%) & 245 (3.48\%) \\

Schizophrenia & 24 (4.93\%) & 82 (3.37\%) & 20 (1.42\%) & 79 (1.12\%) \\

Bipolar Disorder & 175 (35.93\%) & 406 (16.67\%) & 303 (21.52\%) & 679 (9.64\%) \\
\bottomrule
\end{tabular}
\end{table}

\section{Hybrid \& Integrative Learning Framework}\label{sec:model}

\subsection{Setup \& Formulation}\label{sec:model:1}

For the multi-record data, let $\bY = (\bY_p,\bY_s) = (\by_1\trans, \ldots, \by_n\trans)\trans \in\mathbb R^{n\times q}$ be the response matrix of $n$ independent observations on the $q$ outcomes or response variables, where $\bY_p \in\mathbb R^{n\times q_0}$ 
consists of the $q_0$ primary/target outcomes, and $\bY_{s} \in\mathbb R^{n\times (q-q_0)}$ 
consists of the $q-q_0$ surrogate outcomes; let $\bX = (\bx_1\trans, \ldots, \bx_n\trans)\trans \in \mathbb R^{n\times p}$ denote the feature matrix associated with $\bY$, constructed from the historical records. Similarly, for the single-record data, let $\widetilde {\bY} = (\widetilde {\bY}_p, \widetilde {\bY}_s) \in \mathbb R^{n_1\times q}$, with $\widetilde {\bY}_p \in \mathbb R^{n_1\times q_0}$  and  $\widetilde {\bY}_s \in \mathbb R^{n_1\times (q-q_0)}$ denoting the target and surrogate outcomes, respectively. In our motivating application, the primary/target outcome is the occurrence of a suicide attempt ($q_0=1$), and the surrogate outcome consists of the seven concurrent mental health disorders ($q - q_0 = 7$), listed in Table \ref{tab:surrogate_summary}. 


We consider the general case that the multiple outcomes can be of mixed types, e.g., continuously-valued measures such as blood pressure, binary indicators such as occurance of suicide attempt, and counts such as certain virus load. Assume that the $i$th entry of the $k$th outcome, denoted as $y_{ik}$, follows a distribution from the exponential dispersion family, with natural parameter $\theta_{k} \in \mathbb{R}$ and dispersion parameter $\phi_{k} \in \mathbb{R}$. Then the probability density function of $y_{ik}$ could be expressed as
\begin{equation}\label{eq:exp_density}
    f_k(y_{ik};\theta_{ik},\phi_k) = \exp\bigg\{\frac{y_{ik} \theta_{ik}-b_k(\theta_{ik})}{a_k(\phi_k)}+c_k(y_{ik};\phi_k)\bigg\},
\end{equation}
where $a_k(\cdot),b_k(\cdot),c_k(\cdot)$ are determined by the specific distribution of the $k$th outcome; see \citet{LuoDeyChen2015} for some specific examples in the exponential-dispersion family, including Bernoulli, Normal, and Poisson distributions. 
As such, we use $(y_{ik}, \bx_i;\theta_{ik},\phi_k)$, $i=1,\dots,n$ to denote the multi-record data and their associated parameters, and use $(\widetilde y_{ik};\widetilde{\theta}_{ik}, \widetilde{\phi}_{k})$, $i=1,\dots, n_1$ to denote the single-record data and their associated parameters. Accordingly, we define $\bTheta = (\theta_{ik})_{n\times q} = (\btheta_1\trans, \ldots, \btheta_n\trans)\trans \in \mathbb{R}^{n\times q}$ and $\bphi = (\phi_1,\ldots, \phi_q)\trans$ be the natural parameter matrix and dispersion parameter vector for the multi-record data, and $\widetilde \bTheta = (\widetilde\theta_{ik})_{n\times q} = (\widetilde\btheta_1\trans, \ldots, \widetilde\btheta_{n_1}\trans)\trans \in \mathbb{R}^{n_1\times q}$ and $\widetilde\bphi = (\widetilde\phi_1,\ldots, \widetilde\phi_q)\trans$ the counterparts for the single-record data.

For the multi-record data, the main statistical problem is to model a multivariate outcome with a large number of features; such problems are increasingly prevalent across various scientific domains. A fundamental tool is the celebrated reduced-rank regression (RRR), in which predictors and responses are assumed to be interconnected via a small set of latent variables, manifested by the low-rankness of a parameter matrix. Extending this approach, reduced-rank vector generalized linear models \citep{yee2003} and mixed-response reduced-rank regression \citep{luo2018leveraging} have been developed to encompass a broader spectrum of outcome distributions. See \citet{reinsel2022} for a comprehensive account of the reduce-rank methods. 
The idea of low-rank model has also been widely explored across various machine learning techniques. For example, deep neural networks with a limited number of intermediate hidden units, or autoencoder models with a reduced latent dimensionality, can all be viewed as nonlinear generalizations of reduced-rank models \citep{wang2016auto}.

Motivated by the abovementioned works, we shall construct a framework that captures the underlying low-rank structure, connecting the natural parameters to the predictors as well as connecting multi-record and single-record data. For the multi-record data, we link the feature $\X$ to the response $\Y$ through a general encoder-decoder setup. Let $\bm h: \mathbb R^p \to \mathbb R^r$ be an ``encoder'' function, which maps the $p$-dimensional input feature space to an $r$-dimensional intermediate latent representation space with $r\leq \min(p,q)$. Additionally, let $\bm g: \mathbb R^r \to \mathbb R^q$ be a ``decoder'' function that maps the latent space to the $q$-dimensional space of the natural parameters $\theta_{ik}$ of the responses. 
This encoder-decoder setup can be expressed as 
\begin{align}
\btheta_{i}  = (\bm g \circ \bm h)(\bx_i), \qquad i= 1,\ldots, n,\label{eq:encoderdecoder}
\end{align}
where $\circ$ denotes functional composition. Because we set $r \leq \min(p,q)$, this representation inherently exhibits the low-rank structure by construction. For the single-record data, we adopt an autoencoder setup in which the outcomes are directly encoded from the response through an encoder $\widetilde{\bm h}: \mathbb R^q \to \mathbb R^r$. Critically, the latent representation is then mapped to the natural parameters through the same decoder $\bm g$. The autoencoder setup is expressed as 
\begin{align}
\widetilde\btheta_{i}  = (\bm g \circ \widetilde{\bm h})(\by_i), \qquad i= 1,\ldots, n_1,\label{eq:autoencoder}
\end{align}
Here the encoder or decoder functions $\bm h, \bm g, \widetilde{\bm h}$ are within certain classes of functions, i.e., $\bm h \in \mathcal H, \bm g \in \mathcal G$ and $ \widetilde{\bm h} \in \widetilde{\mathcal H}$, respectively.

We term the above general framework, specified in \eqref{eq:encoderdecoder} and \eqref{eq:autoencoder} with the distributional setup in \eqref{eq:exp_density}, as the \textit{Hybrid \& Integrative Reduced Rank Regression} (HiRRR). As illustrated in Figure~\ref{fig:diag_semi_rr}, the shared decoder successfully links the two sets of data, and it implies that the underlying latent representations remain the same, i.e., the single-record and multi-record responses are associated with the same latent variables, no matter whether they are learned through features or the responses themselves.

\subsection{Estimation Criterion} 

To enable the hybrid \& integrative learning, we conduct model estimation by
\begin{equation}\label{eq:mixed_obj_general}
    \begin{split}
    \underset{\bm g \in \mathcal G,\;\bm h\in\mathcal H,\;\widetilde{\bm h}\in\widetilde{\mathcal H},\;\bphi, \widetilde{\bphi}}{\min}
    \quad
    &-\sum_{i=1}^n\sum_{k=1}^{q} w_{ik} l_k(\theta_{ik},\phi_k; y_{ik}, \bx_i)
    -\lambda \sum_{i=1}^n \sum_{k=1}^{q} \widetilde w_{ik} l_k(\widetilde \theta_{ik},\widetilde\phi_k; \widetilde y_{ik})\\
    &\text{s.t. } \theta_{ik}= g_k (\bm h(\bx_i)),\; \widetilde \theta_{ik} = g_k (\widetilde{\bm h}(\by_i)), 
\end{split}
\end{equation}
where $l_k$ is the corresponding log-likelihood function of $f_k$ in~\eqref{eq:exp_density}, and $g_k$ is the $k$th component of the vector valued function $\bm g$. Alternatively, the criterion can be expressed in matrix form as
\begin{equation}\label{eq:mixed_obj_general_matrix}
    \begin{split}
    \underset{\bm g \in \mathcal G,\;\bm h\in\mathcal H,\;\widetilde{\bm h}\in\widetilde{\mathcal H},\;\bphi,\widetilde{\bphi}}{\min}
    \quad
    &L(\bTheta,\widetilde\bTheta, \bphi,\widetilde\bphi) \equiv \left\{ - l_{W}\big(\bm\Theta,\bphi; \bY, \bX \big)
    -\lambda \; l_{\widetilde {W}}\big( \widetilde{\bm\Theta}, \widetilde{\bm\phi}; \widetilde{\bY} \big) \right\}\\
    &\text{s.t. } \bm\Theta = \bm G_n \circ \bm H_n(\bX),\; \widetilde{\bm\Theta} = \bm G_{n_1} \circ \widetilde {\bm H}_{n_1}(\widetilde{\bY}), 
    \end{split}
\end{equation}
where $l_{W}(\cdot)$ is the weighted log-likelihood function with the entrywise weights in $\bW = (w_{ik})$, and $\bm H_n$/$\bm G_n$ are the matrix counterparts of $\bm h$/$\bm g$, which operate on the matrix in a row-wise manner; the notations for the single-record data are similarly defined.

In \eqref{eq:mixed_obj_general} and \eqref{eq:mixed_obj_general_matrix}, we use weighted negative log-likelihood to construct the loss function. The pre-specified weights $w_{ik},\widetilde w_{ik}$ serve to modulate the relative impact of each entry in the model. In particular, the weights could allow the target outcome and the surrogate outcomes to contribute differently to the model's overall parameter estimation. The inclusion of the tuning parameter $\lambda\in[0,1]$ adjusts the influence of the unsupervised component, providing a mechanism to regulate the extent to which information from $\widetilde{\bY}$ is leveraged.

The mappings $\bm h, \bm g, \widetilde{\bm h}$ are within the classes $\mathcal H, \mathcal G$ and $\widetilde{\mathcal H}$, respectively. Their specifications lead to different versions of HiRRR. In particular, if $\mathcal H$ and $\mathcal G$ are families of linear functions, they could be parameterized as $\mathcal H=\{\bm h:\mathbb R^{p}\to \mathbb R^{q}, \bx\mapsto \bx\trans \bA \}$ and $\mathcal G=\{\bm g:\mathbb R^{r}\to \mathbb R^{q}, \bx\mapsto \bx\trans \bB\trans+\bmu,\text{ s.t. } \bB\trans\bB=\bI_r\}$. Here, 
the orthogonal constraint is imposed on $\bB$ for parameter identification. If we impose no restrictions on the class $\widetilde{\mathcal H}$, optimizing $\widetilde h$ in $\widetilde{\mathcal H}$ is equivalent to directly optimizing $\widetilde {\bL}=\widetilde{\bH}_{n_1}(\widetilde \bY)$. Then the model in \eqref{eq:encoderdecoder} and \eqref{eq:autoencoder} become a hybrid linear reduced-rank model, i.e., 
\begin{align}
\bTheta = \1_{n}\bmu + \bX\bA\bB\trans, \qquad
\widetilde \bTheta = \1_{n_1}\bmu + \widetilde \bL\bB\trans,\label{eq:rrr}
\end{align}
and the optimizing problem in~\eqref{eq:mixed_obj_general_matrix} becomes
\begin{equation}\label{eq:mixed_obj_linear_matrix}
 \begin{split}
    \underset{\bA,\;\bB,\;\widetilde{\bL},\;\bphi}{\min}
    \quad
    &- l_{W}(\bm\Theta,\bphi; \bY, \bX)
    -\lambda \; l_{\widetilde {W}}(\widetilde{\bm\Theta},\bphi; \widetilde{\bY}) \\
    &\text{s.t. } \bm\Theta = \1_n \bm\mu\trans + \bX\bA\bB\trans,\quad\widetilde{ \bm\Theta} = \1_{n_1} \bm\mu\trans + \widetilde {\bL} \bB\trans,\quad \bB\trans\bB=\bI_r.
    \end{split}
\end{equation}
Here, we have assumed $\bphi = \widetilde{\bphi}$ for simplicity. 
To make the model even more flexible, we could alternatively set the families $\mathcal H$ and $\mathcal G$ as classes of neural networks. For example, suppose $\mathcal H$ is the family of neural networks with two hidden layers, then it can be parameterized as $\mathcal H = \{\bm h: \bx \mapsto \sigma_2\big( \bW_{h2}\;\sigma_1(\bW_{h1}\bx + \bb_{h1}) + \bb_{h2}\big) \}$, where $\sigma_1,\sigma_2$ are two activation functions applied at each layer, $\bW_{h1},\bW_{h2}$ and $\bb_{h1},\bb_{h2}$ are the weight matrices and bias vectors for each layer. The other encoders and decoders can be specified in a similar fashion. To avoid over-fitting and enhance generalization, additional regularization techniques could be applied when fitting the neural networks, such as dropout, $\ell_2$ regularization, and batch normalization \citep{ioffe2015batch,srivastava2014dropout}.


\section{Computation \& Theoretical Considerations}\label{sec:computation}

\subsection{Computational Algorithms}

The proposed general framework covers various data types and model specifications. Here we mainly discuss algorithms for solving the problem~\eqref{eq:mixed_obj_general} under the reduced-rank setup with mixed outcomes. For similarity, we assume the dispersion parameters are shared; this leaves the relative importance of the supervised and unsupervised components be governed by the tuning parameter $\lambda$, which will be chosen by cross validation.  We also note that for the special case of Gaussian outcomes, the problem admits an explicit solution with no need of an iterative algorithm, which we discuss in Section \ref{sec:theory}.

Recall that we can use $L(\bTheta, \widetilde\bTheta, \bphi) = L(\bA,\bB,\bmu,\widetilde{\bL},\bphi)$ to denote the objective function in~\eqref{eq:mixed_obj_general_matrix}; for simplicity, we may also write $L$ without arguments if no confusion arises. Algorithm~\ref{alg:hirrr_linear} gives a sketch of a blockwise coordinate descent (BCD) algorithm to iteratively update parameters $\bA,\bB,\bmu,\widetilde{\bL},\bphi$ under the reduced-rank setup specified in \eqref{eq:rrr} and \eqref{eq:mixed_obj_linear_matrix}. The updating of $\bA$, $\bmu$, $\widetilde{\bL}$, and $\bphi$ can be accomplished through the gradient descent methods. We have used $\nabla$ to denote the gradient operator; for example, $\nabla_A L$ represents the partial derivative of $L$ with respect to $\bA$. The detailed updating rules related to these parts have been thoroughly discussed in \citet{luo2018leveraging}. 

The distinct challenge in our problem is the updating of $\bB$ under an orthogonality constraint. To address this, at each $(t+1)$th iteration and after updating $\bA$, we construct a quadratic majorizer or a surrogate function with respect to $\bB$, denoted as $L_m$, which satisfies that $L_m(\bB;\bA^{(t+1)},\bB^{(t)},\bmu^{(t)},\widetilde{\bL}^{(t)},\bphi^{(t)}) \geq L(\bA^{(t+1)},\bB^{(t)},\bmu^{(t)},\widetilde{\bL}^{(t)},\bphi^{(t)})$ for any $\bB$ with equality holds when $\bB = \bB^{(t)}$. The optimization $L_m$ with respect to $\bB$ is then reduced to a Procrustes problem \citep{schonemann1966generalized}. The existence of $L_m$, in Algorithm~\ref{alg:hirrr_linear}, is guaranteed by the bounded second derivatives of the log-likelihood for Gaussian or Bernoulli distributions; or by an empirical bound for that of the Poisson distribution. From the algorithmic construction the objective function is non-increasing along the iterations, and the algorithm is stable and always convergent in our numerical studies. 


\begin{algorithm}[htp]
  \caption{HiRRR algorithm under general reduced-rank setting}
  \label{alg:hirrr_linear}
  \begin{algorithmic}
    \STATE {\bf{Parameters}} rank $r$, tuning parameter $\lambda$, weights $\bw,\tilde \bw$, and tolerance level $\epsilon$; $t=0$.
    \STATE {\bf{Input}} $\bX,\bY$ and $\widetilde{\bY}$.
    \STATE {\bf{Initialization}} $\bA^{(0)}, \bB^{(0)}, \bmu^{(0)}, \widetilde{\bL}^{(0)}, \widetilde{\bphi}^{(0)}$.
    \REPEAT
    \STATE $\bA$\textit{-step (gradient descent)}:
    \STATE \quad $\bA^{(t+1)} \gets \bA^{(t)} + t_A\nabla_{A} L$, where $t_A$ is the step size.
    \STATE $\bB$\textit{-step (majorization-minimization and Procrustes problem)}:
    \STATE \quad Construct the quadratic majorizer at current $\bB^{(t)}$ of $L$ as $L_m(\bB;\bB^{(t)})$. 
    \STATE \quad $\bB^{(t+1)} \gets {\arg\min}_{\bB} L_m(\bB;\bB^{(t)})\text{ s.t. }\bB\trans\bB = \bI_r$.
    \STATE $\bmu$\textit{-step (gradient descent)}:
    \STATE \quad $\bmu^{(t+1)} \gets \bmu^{(t)} + t_\mu\nabla_{\mu} L$, where $t_\mu$ is the step size.
    \STATE $\widetilde{\bL}$\textit{-step (gradient descent)}:
    \STATE \quad $\widetilde{\bL}^{(t+1)} \gets \widetilde{\bL}^{(t)} + t_{\tilde L}\nabla_{\tilde L} L$, where $t_L$ is the step size.
    \STATE $\bphi$\textit{-step}:
    \STATE \quad $\bphi^{(t+1)} \gets \arg\min_{\phi}L$.
    \STATE $t\gets t+1$.
    \UNTIL{$|L^{(t+1)} - L^{(t)}| / (|L^{(t)}|+ \epsilon) <\epsilon$}. \\
    \RETURN $\bA, \bB, \bmu, \widetilde{\bL}, \widetilde{\bphi}$.
  \end{algorithmic}
\end{algorithm}

For the case of multivariate binary responses, which covers the motivating application of suicide risk study, we provide implementation details in Algorithm~\ref{alg:hirrr_linear_binomial}. Note that the dispersion parameter is one for Bernoulli distribution, and we have used the notation $\plogis(\theta)=\exp(\theta)/(1+\exp(\theta))$. The derivation is given in Appendix~\ref{sec: solution_glm}. 

\begin{algorithm}[t]
  \caption{HiRRR under reduced-rank setting with binary responses}
  \label{alg:hirrr_linear_binomial}
  \begin{algorithmic}
    \STATE {\bf{Parameters}} rank $r$, tuning parameter $\lambda$, tolerance level $\epsilon$; $t=0$.
    \STATE {\bf{Input}} $\bX,\bY$ and $\widetilde{\bY}$.
    
    \STATE {\bf{Initialization}} $\bmu^{(0)},\bA^{(0)},\bB^{(0)},\tilde{\bL}^{(0)}$.
    \STATE Calculate $(\bX\trans\bX)^{+}$ at first.
    \REPEAT 
    \STATE $\bA$\textit{-step}:
    {
    \STATE \qquad $\bA^{(t+1)} \gets \bA^{(t)} 
    + 4(\bX\trans\bX)^{+}\bX\trans[\bY-\plogis(\bX\bA^{(t)} {\bB^{(t)}}\trans + \1_n \bmu^{(t)} \trans)]\bB^{(t)} $.
    }
    \STATE $\bB$\textit{-step}:
    {
    \STATE \qquad $\bE^* \gets \bX\bA^{(t+1)}\bB^{(t)}\trans+4\big[\bY-\plogis(\bX\bA^{(t+1)}{\bB^{(t)}}\trans + \1_n \bmu^{(t)}\trans)\big]$,
    \STATE \qquad $\widetilde{\bE}^* \gets \widetilde{\bL}^{(t)}{\bB^{(t)}}\trans+4\big[\widetilde{\bY}-\plogis(\widetilde{\bL}^{(t)}{\bB^{(t)}}\trans + \1_{n_1} {\bmu^{(t)}}\trans)\big]$.
    \STATE \qquad Denote the SVD as $\bU\bD\bV\trans \gets \bE^* \trans \bX\bA^{(t+1)} + \lambda \widetilde{\bE}^*\trans\widetilde{\bL}^{(t)}$; 
    \STATE\qquad $\bB^{(t+1)} \gets \bU_r\bV_r\trans$ keeps the first $r$ singular vectors.
    }
    \STATE $\bmu$\textit{-step}:
    {
    \STATE \qquad $\bE^* \gets \bX\bA^{(t+1)}\bB^{(t+1)}\trans+4\big[\bY-\plogis(\bX\bA^{(t+1)}{\bB^{(t+1)}}\trans + \1_n \bmu^{(t)}\trans)\big]$,
    \STATE \qquad $\widetilde{\bE}^* \gets \widetilde{\bL}^{(t)}{\bB^{(t+1)}}\trans+4\big[\widetilde{\bY}-\plogis(\widetilde{\bL}^{(t)}{\bB^{(t+1)}}\trans + \1_{n_1} {\bmu^{(t)}}\trans)\big]$,
    \STATE \qquad $\bmu^{(t+1)} \gets \bmu^{(t)} + (\bE^*\trans\1_n +\lambda\widetilde{\bE}^*\trans\1_{n_1})/(n+\lambda n_1)$.
    }
    \STATE $\widetilde{\bL}$-step:
    {
    \STATE\qquad $\widetilde {\bL}^{(t+1)} \gets \widetilde {\bL}^{(t)} 
    + 4 [\widetilde{\bY}- \plogis(\widetilde {\bL}^{(t)} {\bB^{(t+1)}}\trans + \1_{n_1} {\bmu^{(t+1)}}\trans) ]\bB^{(t+1)}$.
    }
    \STATE $t \gets t+1$
      
    \UNTIL{$|L^{(t+1)} - L^{(t)}| / (|L^{(t)}|+ \epsilon) < \epsilon$}.\\
    \RETURN $\bA, \bB, \bmu, \widetilde{\bL}, \widetilde{\bphi}$.    
  \end{algorithmic}
\end{algorithm}

We also provide a sketch on how we train HiRRR model with $\mathcal H,\mathcal G,\widetilde{\mathcal{H}}$ being neural networks in Appendix~\ref{sec:apd:hirrr_nn}. 
The training objective is the loss function in~\eqref{eq:mixed_obj_general_matrix}, which incorporates both supervised and unsupervised components to jointly optimize predictions across mixed data types. The algorithm leverages back propagation to iteratively adjust network weights of $\bm h$, $\bm g$, and $\widetilde{\bm h}$.

\subsection{Theoretical Considerations}\label{sec:theory}

To better understand how the HiRRR framework could potentially improve statistical estimation, we consider analyzing the following fundamental reduced-rank setup, 
\begin{align*}
    \bY 
     = \bX\bC + \bE  = \bX\bA\bB\trans + \bE, \qquad
    \widetilde{\bY}
     = \widetilde{\bX}\bC + \widetilde{\bE} = \widetilde{\bL}\bB\trans + \widetilde{\bE},  
\end{align*}
where $\bC = \bA\bB\trans$ gives its rank-$r$ decomposition with $\bA \in \mathbb{R}^{p\times r}$ and $\bB\in \mathbb{R}^{q\times r}$, and consequently $\widetilde{\bL} = \widetilde{\bX}\bA$. The error matrices $\bE = (\be_1\trans,\dots, \be_n\trans)\trans \in \mathbb R^{n\times q}$ and $\widetilde{\bE} = (\widetilde\be_1\trans,\dots, \widetilde\be_{n_1}\trans)\trans \in \mathbb R^{n_1\times q}$ are independent, and each row of $\bE$  and $\widetilde{\bE}$ is with mean zero and covariance matrix $\bSigma_e$ and $\widetilde{\bSigma}_{e}$, respectively. 
In this setup, the responses are assumed to be associated with the latent factors through the same loading matrix $\bB$ in both the supervised and unsupervised components. However, the latent factors are realized as $\bX\bA$ in the supervised component but are completely latent as $\widetilde\bL$ in the unsupervised component. 

Consider the unweighted estimation criterion in \eqref{eq:mixed_obj_linear_matrix} with rank $r$; the problem becomes 
\begin{equation}
\min_{\bA,\bB,\widetilde {\bL}}\quad\|\bY - \bX\bA\bB\trans\|_F^2 + \lambda \|\widetilde {\bY}-\widetilde {\bL} \bB\trans\|_F^2, \qquad \mbox{ s.t. } \bB\trans \bB = \bI. \label{eq:HiRRR_TH}
\end{equation}
It can be readily shown that  there is a closed-form solution for $\bC = \bA\bB\trans$. 

\begin{proposition}\label{prop:1}
Define
$\widehat{\bM}
  \;=\;
  {\bY}\trans\,{\bP}_X\,{\bY}
  \;+\;
  \lambda \;\widetilde{{\bY}}\trans\,\widetilde{{\bY}}$,
where $\bP_X = {\bX}\,({\bX}\trans {\bX})^+\,{\bX}\trans$ is the projection matrix onto the column space of $\bX$ with $()^+$ denoting the Moore–Penrose inverse of the enclosed matrix. Let ${\bV_r}$ be the sub-matrix consisting of the top-$r$ eigenvectors of $\widehat{\bM}$.
Then the HiRRR estimator of $\bC$ by solving \eqref{eq:HiRRR_TH} is given by $\widehat {\bC}=\widehat{\bA}\widehat{\bB}\trans$, with 
\begin{align}
\widehat {\bA}=(\bX\trans \bX)^{+}\bX\trans \bY \bV_r, \qquad \widehat{\bB} = \bV_r. 
\end{align}
\end{proposition}

The detailed derivation is given in the Appendix~\ref{sec: solution_lm}. The form of this explicit solution suggests that HiRRR utilizes the additional information from the single-record data to estimate the column space of $\bC$, that is, the estimation of $\bB$, while the estimation of its row space is the same as the conventional reduced-rank regression \citep{reinsel2022}. 

To gain further insight on the performance of HiRRR, we thus focus on developing a non-asymptotic error bound on estimating the shared loading matrix $\bB$. In what follows, we use $\| \cdot \|$ to denote the $\ell_2$ norm of the enclosed vector, and use $\|\cdot\|_{\mathrm{op}}$ and $\mbox{Tr}(\cdot)$ to denote the operator norm and trace of the enclosed square matrix, respectively. For two symmetric matrices, $\bA_1 \succeq \bA_2$ means $\bA_1 - \bA_2$ is nonnegative definite. We use $a \lesssim b$ to denote that $a$ is less than or equal to $b$ up to a multiplicative constant.

From Proposition \ref{prop:1}, we know that the estimated loading matrix $\widehat{\bB}$ is the top‐$r$ eigenvectors of $\widehat{\bM} = {\bY}\trans\,{\bP}_X{\bY} +\lambda \widetilde{{\bY}}\trans\,\widetilde{{\bY}}$. Define the population counterpart of $\widehat{\bM}$ as $\bM_{\mathrm{pop}}$, 
$$
  \bM_{\mathrm{pop}}
  \;=\;
  \mathbb{E}\bigl[{\bY}\trans\,{\bP}_X\,{\bY}\bigr]
  \;+\;
  \lambda\;\mathbb{E}\bigl[\widetilde{{\bY}}\trans\,\widetilde{{\bY}}\bigr].
$$
We first bound the noise fluctuation $\|\widehat{\bM} - \bM_{\mathrm{pop}}\|_{\mathrm{op}}$ using matrix concentration inequalities \citep{Tropp2012}, under a sub-Gaussian assumption. We then show that the true loading matrix ${\bB}^\star$ is the top‐$r$ eigenvectors of a matrix $\bM$ that is very ``close'' to $\bM_{\mathrm{pop}}$, under some reasonable conditions on the signal strength and noise separation. The desired subspace angle bound is then obtained by applying Davis-Kahan $\sin\Theta$ Theorem \citep{YuSamwrth2014}. The main results are summarized as follows. 

\begin{theorem}[Non-Asymptotic Subspace Error]
\label{thm:HiRRR-Subspace}
Consider the HiRRR estimator $\widehat{{\bB}}$ (the top-$r$ eigenvectors of $\widehat{\bM}$) under Assumptions \eqref{as:1} -- \eqref{as:4} presented in Section \ref{sec:proof}. There exist positive constants $c_3, c_4, c_5>0$ 
such that with probability at least 
$1 - c_3/(nq) - c_4/(n_1q) - c_5/(n_1r)$,
it holds that 
$$
  \bigl\|\widehat{{\bB}}\widehat{{\bB}}\trans 
          - {\bB}^\star {\bB}^\star\trans\bigr\|_{\mathrm{op}}
     \;\;\lesssim\;\;
  O\!\Bigl(\,\tfrac{\sqrt{q}}{\sqrt{\,n + \lambda\,n_1\,}}\Bigr).
$$
\end{theorem}

The results in Theoren \ref{thm:HiRRR-Subspace} demonstrate our HiRRR estimator leverages both data sources, treating the single‐record portion as effectively increasing the sample size from $n$ to $n + \lambda\,n_1$; as a result, the subspace estimation error scales on the order of $\sqrt{q}/\sqrt{n + \lambda n_1}$. Also, when $\lambda = 0$ or $n_1 = 0$, we can similarly show that with high probability, $\bigl\|\widehat{{\bB}}\widehat{{\bB}}\trans 
          - {\bB}^\star {\bB}^\star\trans\bigr\|_{\mathrm{op}} \lesssim O(\sqrt{q/n})$. The dimension inflation of $\sqrt{q}$ commonly appears in multi-response operator-norm bounds. Therefore, HiRRR improves over standard RRR, thereby illustrating the potential benefit of incorporating single‐record data.  We remark that $\lambda>0$ is left as a user-chosen weight. In practice, it may depend on the error variances of the two parts and the sample sizes.


\section{Simulation}\label{sec:simulation}

We conducted simulation studies under the generalized reduced-rank regression setups, with either continues or binary outcomes. We mainly compared the generalized linear model (GLM), the reduced-rank regression (RRR) with surrogate responses, and the hybrid \& integrative reduced-rank regression (HiRRR). The GLM method utilizes neither the integrated outcomes or the single-record data, while the RRR method utilizes the latter and the HiRRR method utilizes both. All methods were tuned by 5-fold cross validation. 

The synthetic data were generated as follows. The entries of the covariate matrices $\bX\in\mathbb R^{n\times p}$ and $\widetilde {\bX}\in\mathbb R^{n_1\times p}$ were generated independently from $\mathrm{N}(0, 1)$, the standard normal distribution. The coefficient matrix was generated as $\bC = b\bC_1\bC_2$, where $b$ is a scalar controlling its overall magnitude, and the entries of $\bC_1\in \mathbb R^{p\times r}$, $\bC_2\in \mathbb R^{r\times q}$ were drawn independently from $\mathrm N(0,1)$. Since the primary outcome is supposed to have a stronger association with the covariates than the surrogate outcomes, we permuted the columns of $\C$, such that its first column, denoted as $\bbeta$ and corresponding to the primary outcome, has the largest $\ell_2$ norm among all the columns. For continuous outcomes, we generated the random error matrix $\bE$ with entries from $\mathrm{N}(0, 1)$, and the observed outcomes were then computed as $\bY = \bX \bC + \bE$. For binary outcomes, we set the intercept $\bmu=(\mu_1,0,\dots,0)\trans$, where $\mu_1$ was chosen to control the prevalence of the target outcome. We then computed the probabilities and sampled the entries of $\bY$ and $\widetilde{\bY}$ under the logistic regression setup.

To mimic the real application, we set the sample size of the multi-record data to $n=2000$, and the sample size of the single-record data to $n_1 = 7000$. The number of predictors was set to $p = 300$, and the number of target outcomes was set to $q_0=1$ corresponding to the first column of the outcome matrix. We experimented with varying numbers of outcomes ($q$), true ranks of the coefficient matrices ($r$), and signal strength ($b$) or rarity of the target outcome ($\mu_1$). The settings were summarized as follows.

\begin{enumerate}\label{sim:binary_linear_settings}
    \item[\textbf{Scenario 1}] \label{continuous_linear}
    (Continuous outcomes)
    
    $(q,r)=(10,3), (30,3), (100,10)$; $b = 0.01, 0.05, 0.10, 0.15$.
    \item[\textbf{Scenario 2}] \label{binary_linear}
    (Binary outcomes)
    
    $(q,r)=(10,3), (30,3), (100,10)$; $b = 0.05, 0.10$; prevalence for the target outcome at $10\%$ or $20\%$.
\end{enumerate}

For the evaluation of estimation performance, we considered several metrics from different perspectives. We reported the mean squared errors for estimating the target parameters, Er($\bbeta$), and for estimating the overall coefficient matrix, Er($\bC$). Let the singular value decomposition (SVD) of the true coefficient matrix and the estimated coefficient matrix be $\bC = \bU_C\bD_C\bV_C\trans$ and $\widehat{\bC} = \widehat{\bU}_C\widehat{\bD}_C\widehat{\bV}_C\trans$, respectively. We also examined the discrepancy between the true and estimated row/column spaces and the singular values. These metrics are defined as follows: $\mathrm{Er}(\widehat{\bbeta}) = \|\widehat{\bbeta} - {\bbeta}\|^2/p$, $\mathrm{Er}(\widehat{\bC}) = \|\widehat{\bC} - {\bC}\|_F^2/(pq)$, $\mathrm{Er}(\widehat {\bU}_C) = \|\widehat {\bU}_C\widehat {\bU}_C\trans- {\bU}_C{\bU}_C\trans\|_F^2/r$, $\mathrm{Er}(\widehat {\bV}_C) = \|\widehat {\bV}_C\widehat {\bV}_C\trans- {\bV}_C{\bV}_C\trans\|_F^2/r$, and $\mathrm{Er}(\widehat {\bD}_C) = \|\widehat{\bD}_C - {\bD}_C\|_F^2/r$.
To evaluate the prediction performance, 
we report the mean squared prediction errors for the target outcome and all the outcomes as $\mbox{Pred}(\widehat \bbeta) = \|\bX \bbeta - \bX\widehat{\bbeta}\|^2/n$, and $\mbox{Pred}(\widehat \bC) = \|\bX \bC - \bX\widehat{\bC}\|_F^2/(nq)$, respectively. In the case of binary responses, we additionally report AUC, PRAUC, and  sensitivity/PPV at 90\% or 95\% specificity for the target outcome, based on randomly generated multi-record testing data with sample size $2000$. The simulation under each setting is repeated 100 times, and we report 10\% trimmed means and standard errors of the metrics across all replications.

\begin{table}[htb]
\centering
\scriptsize
\caption{Simulation: Performance metrics for the continuous-outcome scenarios at $q=30$, $r=3$. 
}\label{tab:sim:continuous}
\begin{tabular}{l|rrr|rrr}
\toprule
  & GLM & RRR & HiRRR & GLM & RRR & HiRRR \\
\midrule
& \multicolumn{3}{c}{$b=0.01$} & \multicolumn{3}{c}{$b=0.05$}\\
$\mathrm{Er}(\bU_C)\times10^2$ & 38.01 (3.48) & 37.56 (3.35) & {\bf{37.13}} (3.22) & 1.77 (0.20) & 1.77 (0.20) & {\bf{1.77}} (0.20)\\
$\mathrm{Er}(\bV_C)\times10^2$ & 5.53 (1.07) & 4.36 (0.82) & {\bf{1.67}} (0.35) & 0.17 (0.03) & 0.14 (0.02) & {\bf{0.03}} (0.01)\\
$\mathrm{Er}(\bC)\times10^4$ & 5.89 (0.06) & 0.66 (0.02) & {\bf{0.61}} (0.02) & 5.88 (0.07) & 0.63 (0.02) & {\bf{0.60}} (0.02)\\
$\mathrm{Er}(\bbeta)\times10^4$ & 5.79 (0.36) & 1.73 (0.30) & {\bf{1.70}} (0.28) & 5.92 (0.35) & 1.67 (0.26) & {\bf{1.64}} (0.26)\\
$\mbox{Pred}(\bC)\times10^2$ & 15.02 (0.15) & 1.72 (0.05) & {\bf{1.58}} (0.05) & 14.99 (0.16) & 1.64 (0.05) & {\bf{1.53}} (0.05)\\
$\mbox{Pred}(\bbeta)\times10^2$ & 14.85 (0.91) & 4.45 (0.77) & {\bf{4.35}} (0.70) & 15.13 (0.84) & 4.26 (0.64) & {\bf{4.18}} (0.65)\\
\midrule
& \multicolumn{3}{c}{$b=0.10$} & \multicolumn{3}{c}{$b=0.15$}\\
$\mathrm{Er}(\bU_C)\times10^2$ & 0.45 (0.06) & 0.45 (0.06) & {\bf{0.45}} (0.06) & 0.20 (0.02) & 0.20 (0.02) & {\bf{0.20}} (0.02)\\
$\mathrm{Er}(\bV_C)\times10^2$ & 0.04 (0.01) & 0.03 (0.01) & {\bf{0.01}} (0.00) & 0.02 (0.00) & 0.02 (0.00) & {\bf{0.00}} (0.00)\\
$\mathrm{Er}(\bC)\times10^4$ & 5.89 (0.06) & 0.63 (0.02) & {\bf{0.60}} (0.02) & 5.90 (0.08) & 0.63 (0.02) & {\bf{0.60}} (0.02)\\
$\mathrm{Er}(\bbeta)\times10^4$ & 5.90 (0.31) & 1.68 (0.30) & {\bf{1.66}} (0.29) & 5.82 (0.37) & 1.76 (0.34) & {\bf{1.73}} (0.34)\\
$\mbox{Pred}(\bC)\times10^2$ & 15.03 (0.14) & 1.63 (0.05) & {\bf{1.53}} (0.05) & 15.02 (0.16) & 1.63 (0.05) & {\bf{1.53}} (0.04)\\
$\mbox{Pred}(\bbeta)\times10^2$ & 15.05 (0.77) & 4.31 (0.73) & {\bf{4.24}} (0.72) & 14.89 (0.79) & 4.50 (0.86) & {\bf{4.40}} (0.85)\\
\bottomrule
\end{tabular}

\end{table}

\begin{table}[htb]
\centering
\scriptsize
\caption{Simulation: Performance metrics for the binary-outcome scenarios at $q=10$, $r=3$ with the prevalence of the target outcome at $20\%$. (A few metrics for GLM are not reported as GLM had frequent convergence issues.)
}\label{tab:sim:binary}
\begin{tabular}{l|rrr|rrr}
\toprule
  & GLM & RRR & HiRRR & GLM & RRR & HiRRR\\
\midrule
& \multicolumn{3}{c}{$b=0.05$} & \multicolumn{3}{c}{$b=0.10$} \\
$\mathrm{Er}(\bU_C)\times 10^2$ & 40.48 (6.09) & {\bf{34.11}} (5.24) & 34.18 (5.30) & 33.10 (7.89) & 15.38 (2.47) & {\bf{15.32}} (2.49)\\
$\mathrm{Er}(\bV_C)\times 10^2$ & 5.95 (3.00) & {\bf{1.60}} (0.58) & 2.16 (0.85) & 27.64 (9.36) & 1.31 (0.51) & {\bf{1.17}} (0.36)\\
$\mathrm{Er}(\bC)\times 10^3$ & 7.40 (1.26) & {\bf{1.54}} (0.19) & 1.56 (0.16) & - & 3.57 (0.97) & {\bf{2.90}} (0.47)\\
$\mathrm{Er}(\bbeta)\times 10^3$ & 24.04 (9.18) & 3.75 (1.50) & {\bf{2.95}} (0.86) & - & 10.05 (6.58) & {\bf{4.85}} (1.61)\\
$\mbox{Pred}(\bC)\times 10$ & 19.29 (3.36) & {\bf{3.97}} (0.49) & 4.04 (0.40) & - & 9.56 (2.75) & {\bf{7.64}} (1.33)\\
$\mbox{Pred}(\bbeta)\times 10$ & 63.44 (24.57) & 9.72 (3.90) & {\bf{7.58}} (2.23) & - & 27.27 (18.42) & {\bf{12.58}} (4.24)\\
AUC & 83.00 (2.34) & 86.77 (2.13) & {\bf{86.84}} (2.13) & 90.92 (1.24) & 95.19 (0.95) & {\bf{95.22}} (0.96)\\
PRAUC & 57.21 (4.40) & 64.91 (4.46) & {\bf{65.08}} (4.50) & 73.69 (3.06) & 84.89 (2.64) & {\bf{84.97}} (2.68)\\
Sensitivity (90\% specificity, \%) & 51.24 (5.19) & 60.40 (5.01) & {\bf{60.55}} (5.15) & 70.22 (3.91) & 84.58 (3.51) & {\bf{84.67}} (3.50)\\
Precision (90\% specificity, \%) & 56.00 (2.71) & 60.05 (2.11) & {\bf{60.10}} (2.20) & 63.77 (1.57) & 67.88 (1.11) & {\bf{67.89}} (1.13)\\
Sensitivity (95\% specificity, \%) & 36.55 (4.55) & 45.20 (5.25) & {\bf{45.47}} (5.11) & 55.41 (4.02) & 72.70 (4.48) & {\bf{72.93}} (4.66)\\
Precision (95\% specificity, \%) & 64.38 (3.09) & 69.12 (2.53) & {\bf{69.27}} (2.55) & 73.47 (1.67) & 78.43 (1.12) & {\bf{78.48}} (1.18)\\
\bottomrule
\end{tabular}
\end{table}

Table \ref{tab:sim:continuous} and Table \ref{tab:sim:binary} report the simulation results under Scenario 1 with continuous outcomes and Scenario 2 with binary outcomes, respectively. Across both scenarios, HiRRR consistently outperforms GLM and RRR. In some settings, GLM has had frequent convergence issues. HiRRR shows more substantial improvements in estimating the coefficients, as evidenced by lower errors in \(\mathrm{Er}(\bC)\) and \(\mathrm{Er}(\bbeta)\). This extends to better recovery of the row and column spaces (\(\mathrm{Er}(\bU_C)\) and \(\mathrm{Er}(\bV_C)\)), indicating that HiRRR more effectively uncovers the underlying low-rank structure.  
While HiRRR's prediction improvements over RRR and GLM are evident (\(\mbox{Pred}(\bC)\) and \(\mbox{Pred}(\bbeta)\), as well as classification metrics for binary outcomes), these gains are generally incremental rather than drastic. Nonetheless, HiRRR still yields modest yet consistent enhancements in classification measures such as AUC, PRAUC, sensitivity, and precision at high specificities. Additional simulation results are reported in Appendix \ref{app:sim:rr}.

The noticeable superiority in estimation underscores the advantage of integrating additional surrogate outcomes and single-record data within the reduced-rank framework. By exploiting shared latent structures, HiRRR achieves more accurate coefficient recovery, particularly when the overall signal strength is moderate or low. On the other hand, the incremental improvements in prediction reflect the complexity of translating enhanced estimation into robust out-of-sample performance, especially in large-scale settings where the primary outcome signal may be relatively weak. 

We have also conducted simulation study under the general encoder-decoder framework in Section \ref{sec:model:1}, and detailed settings and results are reported in Appendix \ref{app:sim:nn}. Overall, the HiRRR methods consistently outperforms their RRR counterparts, and both methods could substantially outperform their GLM counterparts. The degree of improvement varied by sample size. In particular, in the smaller dataset \((n,n_1)=(2000,7000)\), the linear HiRRR variant (i.e., without additional neural network layers) achieved the best performance. In contrast, for the larger dataset \((n,n_1)=(20000,70000)\), allowing for nonlinear structures led to much stronger performance gains. The results suggest that neural network–based approaches can better capture complex relationships with sufficient data.

\FloatBarrier
\section{Utilizing Surrogate Responses and Single-Record Data in a Suicide Risk Study}\label{sec:application}

As described in Section \ref{sec:data}, we analyze a case-control cohort of 7975 patients, comprising pediatric, adolescent, and young adult inpatients aged 10–24 years, drawn from the Connecticut HIDD data for the period between October 1, 2012 and September 30, 2017. There were 487 multi-record cases, 2435 multi-record controls, 1408 single-record cases, and 7040 single-record controls. See Figure \ref{fig:hidd_cohort} for a description of the cohort setup, and Tables \ref{tab:hidd_cohort_demo}- \ref{tab:surrogate_summary} for the demographic compositions of the study cohort and the prevalence of seven categories of mental health disorders.

Our goal is to build a statistical model for studying the occurrence of suicide attempts based on historical medical records. For patients with multiple records, historical information was aggregated from the initial record to the penultimate record, with the last record providing the outcome. Predictor variables included demographic characteristics (e.g., age, gender, and race) and diagnosis codes from each prior medical encounter. We converted all ICD-10 codes to their ICD-9 counterparts,  retained the first three digits, and kept the three-digit codes with a prevalence exceeding 0.5\%. Following these steps, a total of $222$ distinct diagnostic conditions remained. In addition, prior mental disorders were also identified based on historical diagnoses and included as covariates for the multi-record dataset.

We considered several competing methods that used only the multi-record data: GLM0: a baseline logistic regression model with suicide attempt as the binary response and only the demographic variables as predictors; GLM: a logistic regression model with suicide attempt as the binary response and both the demographic and the ICD variables as predictors; as the method may encounter convergence issues with rare features, we adopted a marginal screening procedure with Fisher's exact test on the training data to select the top 100 ICD variables.  
RRR: a generalized reduced-rank regression model with suicide attempt and concurrent mental disorders as multivariate response and both the demographic and the ICD variables as predictors; and RRR-NN: the RRR model with both the encoder and decoder fitted as neural networks. We then applied HiRRR (with equal weights for simplicity) and its neural network version HiRRR-NN to leverage the information from both multi-record and single-record data. The rank and the tuning parameters $\lambda$ were selected by 5-fold cross validation. For HiRRR-NN and RRR-NN, we defined the encoders and the decoder as neural networks with 1--4 hidden layers, each with up to 100 hidden units. To prevent over-fitting, the $\ell_2$ regularization was also applied. To facilitate convergence, the models were trained for 200 epochs with a learning rate scheduler: it starts at $0.01$ and reduces by a factor of 0.6 every 50 epochs.

\subsection{Out-of-Sample Prediction Performance}

To assess the performance of various models on predicting suicide for multi-record patients, we randomly partitioned the multi-record patient cohort into a 90\% training set and a 10\% test set. Subsequently, different models were fitted on the training set and evaluated on the testing set. With the HiRRR method, the models were trained using not only the 90\% subset of multi-record data but also the entirety of the single-record data. The out-of-sample performance metrics included the area under the receiver operating characteristic curve (AUC), the area under the precision-recall receiver operating characteristic curves (PRAUC), sensitivity at 90/95\% specificity, and positive predictive value (PPV) at 90/95\% specificity. This random-splitting process was repeated 10 times, and the results were averaged.

\begin{table}[htp]
\centering
\scriptsize
\caption{Application: Performance metrics for various models with the random-splitting procedures.}
\label{tab:app:metrics}
\begin{tabular}{lllllll}
\toprule
\multirow{2}{*}{Model} & \multirow{2}{*}{AUC (\%)} & \multirow{2}{*}{PRAUC (\%)} &\multicolumn{2}{c}{90\% Specificity} & \multicolumn{2}{c}{95\% Specificity}\\
& & & \makecell[c]{Sensitivity (\%)} & \makecell[c]{PPV (\%)} & \makecell[c]{Sensitivity (\%)} & \makecell[c]{PPV (\%)}\\
\midrule
GLM0 & ${53.44\; (4.83)}$ & ${14.92\; (2.46)}$ & ${11.98\; (8.34)}$ & ${18.17\; (10.05)}$ & ${5.78\; (4.76)}$ & ${16.74\; (13.01)}$\\
GLM & ${73.64\; (2.82)}$ & ${32.60\; (4.93)}$ & ${30.31\; (9.61)}$ & ${37.45\; (7.69)}$ & ${13.53\; (4.30)}$ & ${34.88\; (7.81)}$\\
RRR & $\mathbf{75.72\; (2.99)}$ & ${33.98\; (5.22)}$ & ${27.42\; (5.61)}$ & ${35.71\; (6.22)}$ & ${16.18\; (7.98)}$ & ${37.60\; (11.04)}$\\
RRR-NN & ${75.20\; (2.45)}$ & ${32.30\; (4.48)}$ & ${27.35\; (6.93)}$ & ${35.43\; (6.53)}$ & ${14.28\; (5.08)}$ & ${35.83\; (7.90)}$\\ 
HiRRR & ${75.11\; (2.65)}$ & $\mathbf{34.91\; (3.95)}$ & $\mathbf{33.62\; (5.70)}$ & $\mathbf{40.54\; (4.84)}$ & $\mathbf{19.85\; (6.30)}$ & $\mathbf{43.87\; (7.60)}$\\
HiRRR-NN & ${75.67\; (2.63)}$ & ${33.35\; (5.09)}$ & ${29.46\; (6.92)}$ & ${37.22\; (6.76)}$ & ${15.65\; (5.45)}$ & ${38.09\; (7.87)}$\\
\bottomrule
\end{tabular}

\end{table}

\begin{figure}[t]
    \centering
     \begin{subfigure}[t]{0.45\textwidth}
        \centering
        \includegraphics[width=\textwidth]{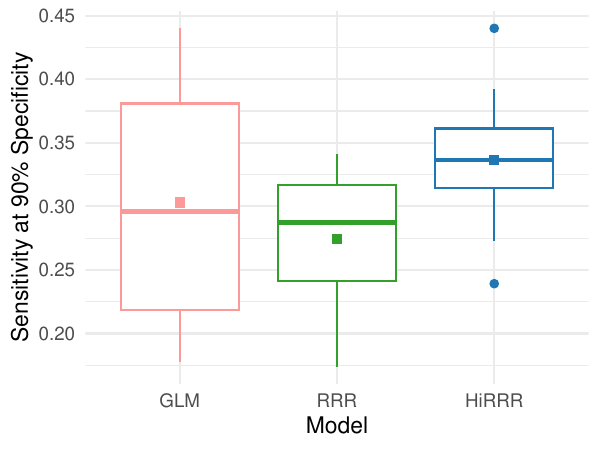}
        \caption{Sensitivity at 90\% Specificity}
    \end{subfigure}
    ~
    \begin{subfigure}[t]{0.45\textwidth}
        \centering
        \includegraphics[width=\textwidth]{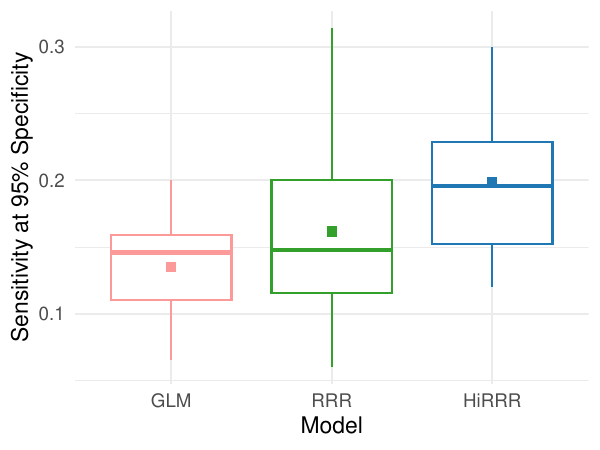}
        \caption{Sensitivity at 95\% Specificity}
    \end{subfigure}%
    
    \begin{subfigure}[t]{0.45\textwidth}
        \centering
        \includegraphics[width=\textwidth]{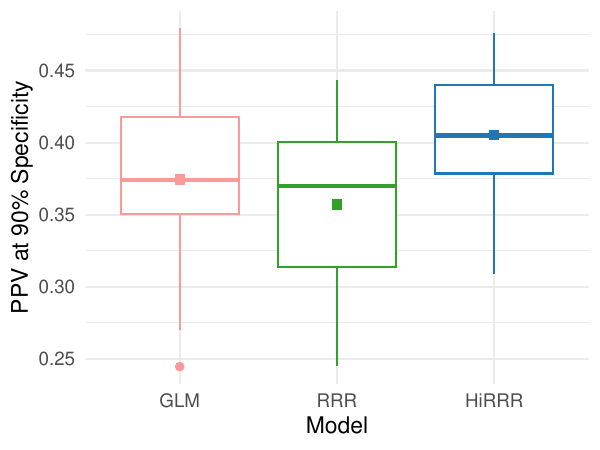}
        \caption{PPV at 90\% Specificity}
    \end{subfigure}%
    ~ 
    \begin{subfigure}[t]{0.45\textwidth}
        \centering
        \includegraphics[width=\textwidth]{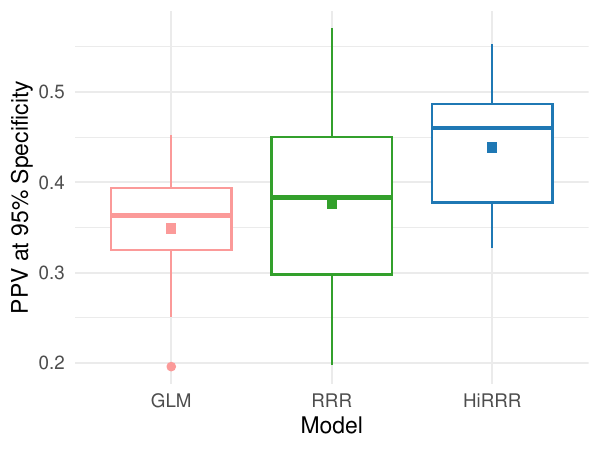}
        \caption{PPV at 95\% Specificity}
    \end{subfigure}
    
    \caption{Application: Boxplots of the performance metrics from the random splitting procedure.} 
    \label{fig:app:metrics}
\end{figure}

Table~\ref{tab:app:metrics} reports the results on the prediction performance from the random splitting procedure, and Figure \ref{fig:app:metrics} provides the boxplots of some evaluation metrics. 
Both RRR and HiRRR selected rank-3 models. GLM models performed the worst. By utilizing multivariate responses, RRR achieved better performance than GLM. The RRR, HiRRR and their neural networks versions had comparable AUC and PRAUC values. Here in this problem, neural network models do not show much improvement comparing to linear reduced-rank models.
Overall, HiRRR performed substantially better in capturing high risk patients than all other models. In particular, comparing HiRRR to GLM and RRR, the sensitivity under 95\% specificity improved from 13.5\% and 16.2\% to 19.9\%, an 47.4\% and 22.8\% improvement, respectively. That is, if we were to select 5\% of patient for further suicide prevention efforts, if successful, using HiRRR could potentially prevent 20\% more suicide attempt cases than using RRR and 47.5\% more cases than using GLM. These results demonstrated the great potential on utilizing surrogate responses and single-record data to inform and improve clinical practice.

\subsection{Estimation \& Selected Risk Factors}

Table \ref{tab:app:variable} presents the top 10 risk factors with the largest averaged standardized coefficients from GLM, RRR, and HiRRR over 10 random splits. We exclude the NN models in this comparison, as they did not substantially improve prediction and require different approaches for feature importance. Both RRR and HiRRR consistently identified key mental health diagnoses, such as \textit{depressive disorders, mood disorders, anxiety, posttraumatic stress disorder, and other psychosocial circumstances}, as the most important risk factors for suicidal behavior. Additional notable predictors included \textit{drug dependence, poisoning by analgesics, and certain adverse effects.} Despite these shared findings, RRR and HiRRR differed in effect size estimates. For instance, the coefficient for depressive disorder was 0.121 in RRR versus 0.373 in HiRRR, suggesting that incorporating single-record data in HiRRR can amplify the relative importance of critical risk factors. In contrast, GLM omitted some mental health–related diagnoses and instead highlighted conditions like alcohol use disorder and disorders of fluid electrolyte and acid-base balance; the latter's link to suicide is less evident clinically. 
Overall, the findings appear clinically coherent, with HiRRR's integration of surrogates and single-record data yielding relatively larger coefficients for key mental health conditions and thereby offering a potentially richer understanding of suicide risk factors.

\begin{table}[htp]
\centering
\tiny
\caption{Application: Top 10 risk factors from fitting GLM, RRR, and HiRRR over 10 random splits. Also reported are the prevalence of each condition/exposure and the log odds ratio between the exposure and the outcome of suicide attempt.}
\label{tab:app:variable}
\begin{tabular}{l L{3.6cm} r r r r r}
\toprule
\makecell[l]{Variable} & \makecell[l]{Description} & \makecell[c]{Prevalence in\\case/control} & \makecell[c]{Log\\odds\\ratio} & \makecell[c]{HiRRR} & \makecell[c]{RRR} & \makecell[c]{GLM}\\
\midrule
Depressive & Major depressive disorder & 60.6\% / 27.3\% & 1.144 & {\bf{0.373}} (0.019) & {\bf {0.121}} (0.017) & {\bf{0.112}} (0.049)\\
ICD-9 296 & Episodic mood disorders & 68.8\% / 31.7\% & 1.285 & {\bf{0.257}} (0.018) & {\bf {0.370}} (0.012) & {\bf{0.343}} (0.034)\\
ICD-9 311 & Depressive disorder, not elsewhere classified & 26.3\% / 14.0\% & 0.623 & {\bf{0.209}} (0.011) & {\bf {0.169}} (0.007) & {\bf{0.173}} (0.015)\\
Posttraumatic & Posttraumatic stress disorder & 17.5\% / 7.1\% & 0.781 & {\bf{0.179}} (0.027) & {\bf {0.105}} (0.016) & -0.008 (0.030)\\
ICD-9 300 & Anxiety, dissociative and somatoform disorders & 41.9\% / 24.0\% & 0.667 & {\bf{0.170}} (0.030) & {\bf {0.160}} (0.018) & {\bf{0.170}} (0.028)\\
ICD-9 528 & Diseases of the oral soft tissues excluding lesions specific for gingiva and tongue & 0.0\% / 1.8\% & -Inf & {\bf{0.151}} (0.031) & 0.042 (0.016) & -1.236 (0.068)\\
ICD-9 965 & Poisoning by analgesics antipyretics and antirheumatics & 3.3\% / 1.0\% & 0.870 & {\bf{0.134}} (0.015) & 0.057 (0.014) & {\bf{0.162}} (0.028)\\
ICD-9 V62 & Other psychosocial circumstances & 44.4\% / 16.4\% & 1.093 & {\bf{0.121}} (0.016) & {\bf {0.104}} (0.015) & {\bf{0.236}} (0.013)\\
ICD-9 304 & Drug dependence & 11.5\% / 4.7\% & 0.737 & {\bf{0.112}} (0.013) & {\bf {0.069}} (0.011) & {\bf{0.177}} (0.023)\\
ICD-9 995 & Certain adverse effects not elsewhere classified & 1.8\% / 5.2\% & -0.945 & {\bf{0.110}} (0.023) & 0.037 (0.014) & 0.056 (0.051)\\
ICD-9 305 & Nondependent abuse of drugs & 32.6\% / 19.2\% & 0.575 & 0.089 (0.019) & {\bf {0.133}} (0.018) & {\bf{0.098}} (0.021)\\
ICD-9 309 & Adjustment reaction & 19.5\% / 9.0\% & 0.703 & 0.016 (0.021) & {\bf {0.082}} (0.017) & 0.041 (0.018)\\
ICD-9 401 & Essential hypertension & 1.8\% / 2.5\% & -0.265 & 0.080 (0.007) & {\bf {0.064}} (0.010) & 0.000 (0.000)\\
ICD-9 276 & Disorders of fluid electrolyte and acid-base balance & 9.0\% / 14.7\% & -0.471 & -0.037 (0.023) & -0.027 (0.015) & {\bf{0.120}} (0.042)\\
Alcohol & Alcohol use disorder & 12.7\% / 5.3\% & 0.741 & 0.092 (0.017) & 0.048 (0.010) & {\bf{0.112}} (0.020)\\
\bottomrule
\end{tabular}

\end{table}

We have also reported the top 10 protective factors identified from GLM, RRR, and HiRRR in Table \ref{tab:app:protective} in Appendix \ref{app:suicide}. Consistent with expectations, all three approaches tended to select relatively rare conditions, many of which had zero prevalence among individuals who attempted suicide or exhibited strongly negative log-odds ratios. Notably, the magnitudes of HiRRR coefficients generally lied between those of RRR and GLM. The reduced-rank structure in RRR can induce strong shrinkage when certain conditions are rare or absent in one group, thereby reducing extreme parameter values but potentially underestimating effect sizes. In contrast, GLM, especially in the presence of quasi-complete separation—may allows coefficients to grow unbounded and even lead to convergence issues. 
With respect to the specific conditions flagged by HiRRR, many appear to be congenital anomalies or procedural complications with low prevalence overall. Further clinical investigation would be required to validate whether and how these conditions or their associated care pathways confer protective effects.

\subsection{Understanding Uniquely Identified Cases by HiRRR}

The HiRRR model leads to a substantial improvement in identifying patients who have attempted suicide. Through our out-of-sample random splitting procedure, we found that when we used either RRR or HiRRR to label the top 10\% of high-risk patients, both methods correctly identified 84 patients with suicide attempts. Additionally, HiRRR uniquely identified another 42 patients with suicide attempts. 

\begin{table}[htp]
\scriptsize
\centering
\caption{Application: Comparison of cases identified by RRR and HiRRR models. }
\label{tab:app:top10_rr_ssrr_comparision}
\begin{tabular}{lL{3.5cm}ccccc}
\toprule
\makecell[l]{Variable} & \makecell[c]{Description} & \makecell[c]{Both RRR/HiRRR\\correct cases\\(exposed/\\nonexposed)} & \makecell[c]{Only HiRRR\\correct cases\\(exposed/\\nonexposed)} & \makecell[c]{P value} & \makecell[c]{Adj. P value}\\
\midrule
Bipolar & Bipolar disorder & 41/43 & 0/42 & 0.000 & 0.000\\
Posttraumatic & Posttraumatic stress disorder & 51/33 & 6/36 & 0.000 & 0.000\\
ICD-9 309 & Adjustment reaction & 51/33 & 6/36 & 0.000 & 0.000\\
Alcohol & Alcohol use disorder & 19/65 & 2/40 & 0.011 & 0.630\\
Drug & Drug use disorder & 33/51 & 7/35 & 0.014 & 0.650\\
ICD-9 305 & Nondependent abuse of drugs & 37/47 & 9/33 & 0.018 & 0.687\\
ICD-9 780 & General symptoms & 7/77 & 10/32 & 0.025 & 0.831\\
ICD-9 338 & Pain, not elsewhere classified & 0/84 & 3/39 & 0.035 & 0.931\\
ICD-9 V15 & Other personal history presenting hazards to health & 30/54 & 7/35 & 0.037 & 0.931\\
ICD-9 296 & Episodic mood disorders & 82/2 & 37/5 & 0.041 & 0.931\\
\bottomrule
\end{tabular}

\end{table}

To understand the unique contributions from the single-record data, we used Fisher’s exact tests to compare the ``both RRR/HiRRR correct'' and ``only HiRRR correct'' groups of patients with suicide attempts, exploring how they differ in terms of historical diagnoses. The $p$-values were adjusted using the Benjamini-Hochberg (BH) procedure. Since the results are exploratory, we have reported all potentially distinctive features based on unadjusted $p$-values with a significance level of 5\% in Table \ref{tab:app:top10_rr_ssrr_comparision}. Analysis of the results shows that the patients correctly identified only by HiRRR had significantly fewer historical diagnoses of certain mental health conditions compared to those identified by both RRR and HiRRR. Specifically, conditions such as bipolar disorder, posttraumatic stress disorder, and substance use disorders were less prevalent among the patients uniquely identified by HiRRR. This indicates that HiRRR effectively identifies at-risk individuals who may lack extensive historical medical records, demonstrating the model's ability to utilize single-record data to enhance the detection of suicide risk.

\FloatBarrier


\section{Discussion}\label{sec:discussion}

We have introduced a novel hybrid and integrative learning framework designed to capitalize on both surrogate responses and single-record data. By anchoring on a shared latent variable structure across multiple outcomes, this framework effectively exploits co-occurring diagnoses and individuals with only a single recorded encounter to bolster model estimation and prediction. Our motivating application on suicide risk demonstrates that the proposed Hybrid \& Integrative Reduced-Rank Regression (HiRRR) model provides notable gains over traditional methods.

There are several potential avenues for future exploration. First, the framework could be extended by considering feature selection. This can be achieved by introducing a row-wise penalty on matrix $\bA$, so that the resulting model aligns with sparse reduced-rank regression \citep{chen2012sparse,chen2012jrssb}. Viewing this sparse structure in the general encoder-decoder setup, it amounts to prune certain links from unimportant or redundant features to the latent representations. Second, extensions to survival analysis may offer a more natural modeling framework for health outcomes such as hospital readmission or mortality, where event times and censoring must be considered. Third, dynamic or longitudinal extensions could capture patient trajectories over time, particularly valuable in healthcare settings where repeated measures are the norm. Last but not the least, domain adaptation or federated learning frameworks might help generalize the hybrid and integrative methods across different hospital systems or large-scale health networks without sharing sensitive patient-level information.

\bibliography{kun-bibtex}{}

\clearpage
\appendix 

\noindent {\bf \LARGE Appendix}

\section{Identification of Major Mental Disorders}\label{sec:apd:mental_health}
\begin{table}[htp]
\scriptsize
\centering
\caption{ICD codes for mental disorders}
\begin{tabular}{ll}
\toprule
Mental Disorders & ICD-9 Codes\\
\midrule
Major Depressive Disorder & 293.83, 296.2, 296.3, 296.9, 298.0, 300.4, 301.12, 309.0\\
Alcohol Use Disorder & 291.0-5, 291.8-9, 303.0-303.9, 305.0, 357.5, 425.5, 571.0-3, 535.3, V11.3\\
Drug Use Disorder & 292.0-1, 304.0-304.9, 305.2-305.8\\
Anxiety Disorder & 300.0, 300.1, 300.2, 799.2\\
Posttraumatic Stress Disorder & 309.81\\
Schizophrenia & 295.0-295.9, V11.0\\
Bipolar disorder & 296.0, 296.1, 296.4-7, 296.80, 296.81, 296.82, 296.89, 296.90, 296.99, V11.1\\
\bottomrule
\end{tabular}
\end{table}

\section{Details of Computation}

\subsection{Details of the Gradients used in HiRRR with Multivariate Bernoulli Outcomes}\label{sec: solution_glm}

Recall $\bTheta = \1_n\bmu\trans + \bX\bA\bB\trans$ and $\widetilde{\bTheta} = \1_{n_1}\bmu\trans + \widetilde{\bL}\bB\trans$, and $L(\bTheta,\widetilde{\bTheta}) = \sum_{i,j}\bigl[y_{ij}\theta_{ij}-\log\bigl(1+\exp(1+\theta_{ij})\bigr)\bigr] + \lambda \sum_{i,j}\bigl[\tilde y_{ij}\tilde\theta_{ij}-\log\bigl(1+\exp(1+\tilde\theta_{ij})\bigr)\bigr]$, then ${\partial L}/{\partial \bTheta} = \bY - {\exp(\bTheta)}/\bigl({1+\exp(\bTheta)}\bigr)$,
and 
\begin{align*}
\frac{\partial L}{\partial \A} &= \bX\trans \frac{\partial L}{\partial \bTheta} \bB =\bX\trans \bY\bB - \bX\trans\frac{\exp(\bTheta)}{1+\exp(\bTheta)}\bB,\\
\frac{\partial L}{\partial \widetilde{\bL}} &= \frac{\partial L}{\partial \widetilde{\bTheta}} \bB =\widetilde{\bY}\bB - \frac{\exp(\widetilde{\bTheta})}{1+\exp(\widetilde{\bTheta})}\bB,\\
\frac{\partial L}{\partial \bmu} &= \Bigl(\frac{\partial L}{\partial \bTheta} \Bigr)\trans \1_{n} + \lambda \Bigl(\frac{\partial L}{\partial \widetilde{\bTheta}}\Bigr)\trans \1_{n_1} = \Bigl[\bY - \frac{\exp(\bTheta)}{1+\exp(\bTheta)}\Bigr]\trans\1_n + \lambda\Bigl[\widetilde{\bY} - \frac{\exp(\widetilde{\bTheta})}{1+\exp(\widetilde{\bTheta})}\Bigr]\trans\1_{n_1} 
\end{align*}
For the univariate function $l(\theta)=y\theta-\log(1+e^\theta)$, there exists a quadratic surrogate function at $\theta_0$, $l(\theta;\theta_0)=-\bigl\{\theta-\theta_0-4\bigl[y-e^\theta/(1+e^\theta)\bigr]\big\}/8$. Therefore, there exists a surrogate function that majorizes $L$ associated with $\bB$ at $\bB_0$ as
\begin{equation*}
    L_m(\bB;\bB_0)=\|\bE^* - \bX\bA\bB\trans\|_F^2 + \lambda \|\widetilde{\bE}^* - \widetilde{\bL}\bB\trans\|_F^2,
\end{equation*}
where $\bE^*=\bX\bA\bB_0\trans + 4\bigl[\bY - \exp(\bTheta)/\bigl(1+\exp(\bTheta)\bigr)\bigr]$ and $\widetilde{\bE}^*=\widetilde{\bL}\bB_0\trans + 4\bigl[\widetilde{\bY} - \exp(\widetilde{\bTheta})/\bigl(1+\exp(\widetilde{\bTheta})\bigr)\bigr]$. Then this surrogate function $L_m(\bB;\bB_0)$ with orthogonal constraint could be solved as a Procrustes problem by 
\begin{align*}
\widehat \bB =\bU \bV\trans, 
\end{align*}
where $\bU,\bV$ corresponds to the singular value decomposition of ${\bE^*}\trans\bX\bA + \lambda{\widetilde{\bE}^*}\trans\widetilde{\bL}=\bU\bD\bV\trans$.

\subsection{HiRRR with Encoder-Decoder}\label{sec:apd:hirrr_nn}

Algorithm \ref{alg:hirrr_nn} gives a sketch of  the model training of HiRRR with neural networks under the general encoder-decoder framework.

\begin{algorithm}[htb]
  \caption{HiRRR with Neural Networks} 
  \label{alg:hirrr_nn}
  \begin{algorithmic}
    \STATE {\bf{Parameters}} structures of $\mathcal H,\mathcal G, \widetilde{\mathcal H}$ (number of hidden layers, number of hidden units, activation functions, dropout rates, etc.), loss function, 
    tuning parameter $\lambda$, number of epoch, optimizer, early stopping condition (if applicable).\\
    \STATE {\bf{Input}} $\bX,\bY$,  $\widetilde{\bY}$ and $\bX_v,\bY_v$ (if early stopping is set).\\
    \STATE {\bf{Initialization}} Initialize weights and biases of the network $\bm h\in\mathcal H,\bm g\in\mathcal G$, and $\widetilde{\bm h}\in \widetilde{\mathcal H} $. 
    \STATE Shuffle the data
    \REPEAT
    \FOR{each epoch}
        \FOR{each mini-batch $mb$}
        \STATE (\textit{Forward Pass}) Calculate $\bTheta_{mb} = \bm G \circ \bm H(\bX_{mb})$ and $\widetilde{\bTheta}_{mb} = \bm G \circ \widetilde{\bm H}(\widetilde{\bY}_{mb})$.
        \STATE (\textit{Loss Calculation}) Calculate loss 
        $-l\big(\bY_{mb}; \bm\Theta_{mb}\big)-\lambda \; l\big(\widetilde{\bY}_{mb}; \widetilde{\bTheta}_{mb})$
        \STATE (\textit{Backward Propagation}) Calculate the gradient of the loss with respect to each parameter
        \STATE (\textit{Update Parameters}) Update the weights and biases parameters with the optimizer
        \ENDFOR
        \STATE Calculate the metrics for validation based on $\bm G \circ \bm H(\bX_v)$ and $\bY_v$.
        \STATE Update early stopping object
    \ENDFOR
    \UNTIL{maximum number of epochs reached or early stopping condition reached.}\\
  \end{algorithmic}
\end{algorithm}

\clearpage

\section{Additional Simulation Results}\label{app:sim}

\subsection{Simulation under Low-Rank Model Setups}\label{app:sim:rr}

Table~\ref{append:tab:gaussian_q10_n17000} and Table~\ref{append:tab:gaussian_q100_n17000} report the performance metrics for the continuous-outcome scenarios with $q=10$ and $q=100$. The results are consistent with those reported in the main paper.  

\begin{figure}[htp]
    \centering
     \begin{subfigure}[t]{0.45\textwidth}
        \centering
        \includegraphics[width=\textwidth]{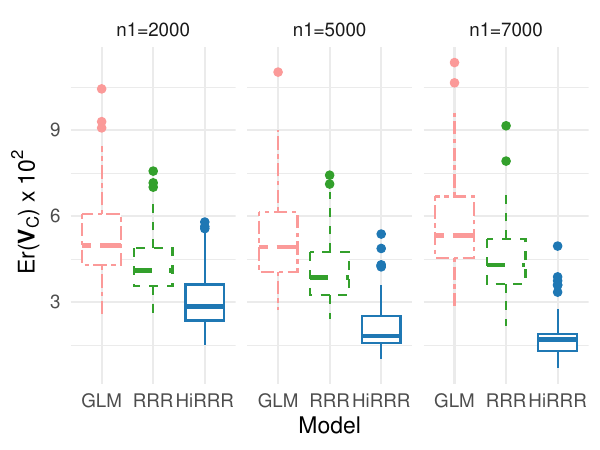}
    \end{subfigure}
    ~
    \begin{subfigure}[t]{0.45\textwidth}
        \centering
        \includegraphics[width=\textwidth]{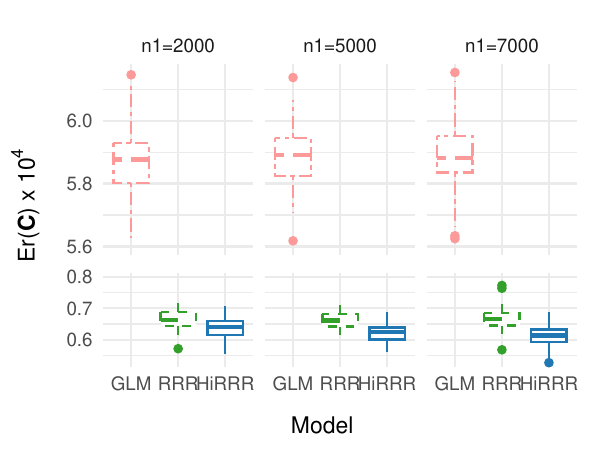}
    \end{subfigure}
    \caption{Simulation: Estimation performance of $\mathrm{Er}(\bV_C)$ and $\mathrm{Er}(\bC)$ at $q=30, r=10$, and $b=0.01$ for continuous-outcome scenarios.}
    \label{append:fig:gaussian_estimation_varying_n1}
\end{figure}

\begin{figure}[htp]
    \centering
     \begin{subfigure}[t]{0.45\textwidth}
        \centering
        \includegraphics[width=\textwidth]{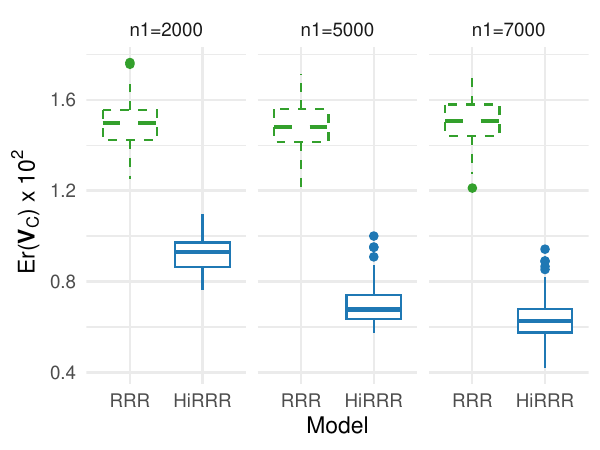}
    \end{subfigure}
    ~
    \begin{subfigure}[t]{0.45\textwidth}
        \centering
        \includegraphics[width=\textwidth]{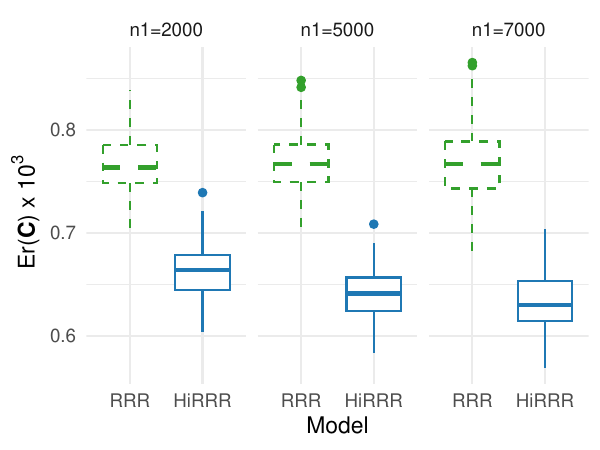}
    \end{subfigure}
    \caption{Simulation: Estimation performance of $\mathrm{Er}(\bV_C)$ and $\mathrm{Er}(\bC)$ at $q=100, r=10$, and $b=0.05$ for binary-outcome scenarios (GLM results were not presented as they fell outside the range).
    } \label{append:fig:binary_estimation_varying_n1}
\end{figure}

Besides, Figure~\ref{append:fig:gaussian_estimation_varying_n1} and Figure~\ref{append:fig:binary_estimation_varying_n1} show the estimation performance of the HiRRR model under varying $n_1$ values of $2000$, $5000$, and $7000$, for the continuous-outcome scenarios and binary outcome scenarios, respectively. These provide empirical evidence regarding the error rate of the HiRRR model. As $n_1$ increases, $\mathrm{Er}(\bV_C)$ appears to decrease in the order of $1/(n+n_1)$. 

\begin{table}[htbp]
\centering
\scriptsize
\caption{Simulation: Performance metrics for continuous-outcome scenarios at $q=10$, $r=3$, and $n_1=7000$. 
}
\label{append:tab:gaussian_q10_n17000}
\begin{tabular}{l|rrr|rrr}
\toprule
  & GLM & RRR & HiRRR & GLM & RRR & HiRRR \\
\midrule
  & \multicolumn{3}{c}{$b=0.01$}  & \multicolumn{3}{c}{$b=0.05$} \\
$\mathrm{Er}(\bU_C)\times10^2$ & 87.58 (8.91) & 86.54 (8.60) & {\bf{85.78}} (8.53) & 6.83 (1.45) & 6.82 (1.45) & {\bf{6.82}} (1.44)\\
$\mathrm{Er}(\bV_C)\times10^2$ & 9.68 (4.97) & 7.31 (3.68) & {\bf{4.16}} (2.27) & 0.18 (0.08) & 0.14 (0.05) & {\bf{0.04}} (0.02)\\
$\mathrm{Er}(\bC)\times10^4$ & 5.87 (0.10) & 1.89 (0.07) & {\bf{1.81}} (0.06) & 5.88 (0.11) & 1.81 (0.05) & {\bf{1.78}} (0.05)\\
$\mathrm{Er}(\bbeta)\times10^4$ & 5.83 (0.33) & 3.37 (0.63) & {\bf{3.25}} (0.63) & 5.93 (0.40) & 3.42 (0.58) & {\bf{3.41}} (0.59)\\
$\mbox{Pred}(\bC)\times10^2$ & 14.96 (0.22) & 4.85 (0.16) & {\bf{4.64}} (0.14) & 14.97 (0.25) & 4.61 (0.13) & {\bf{4.53}} (0.13)\\
$\mbox{Pred}(\bbeta)\times10^2$ & 14.90 (0.77) & 8.66 (1.57) & {\bf{8.34}} (1.54) & 15.05 (0.89) & 8.73 (1.46) & {\bf{8.69}} (1.48)\\
\midrule
  & \multicolumn{3}{c}{$b=0.10$}  & \multicolumn{3}{c}{$b=0.15$} \\
$\mathrm{Er}(\bU_C)\times10^2$ & 1.85 (0.45) & 1.85 (0.45) & {\bf{1.85}} (0.45) & 0.85 (0.23) & {\bf{0.85}} (0.23) & 0.85 (0.23)\\
$\mathrm{Er}(\bV_C)\times10^2$ & 0.05 (0.02) & 0.04 (0.01) & {\bf{0.01}} (0.00) & 0.02 (0.01) & 0.02 (0.01) & {\bf{0.00}} (0.00)\\
$\mathrm{Er}(\bC)\times10^4$ & 5.91 (0.12) & 1.82 (0.05) & {\bf{1.79}} (0.05) & 5.88 (0.10) & 1.81 (0.05) & {\bf{1.78}} (0.06)\\
$\mathrm{Er}(\bbeta)\times10^4$ & 5.98 (0.33) & 3.55 (0.63) & {\bf{3.54}} (0.63) & 5.82 (0.38) & 3.36 (0.54) & {\bf{3.34}} (0.55)\\
$\mbox{Pred}(\bC)\times10^2$ & 15.06 (0.30) & 4.62 (0.13) & {\bf{4.55}} (0.13) & 15.00 (0.25) & 4.60 (0.13) & {\bf{4.53}} (0.13)\\
$\mbox{Pred}(\bbeta)\times10^2$ & 15.14 (0.84) & 9.00 (1.57) & {\bf{8.98}} (1.57) & 14.83 (0.90) & 8.52 (1.32) & {\bf{8.47}} (1.34)\\
\bottomrule
\end{tabular}
\end{table}

\begin{table}[htbp]
\centering
\scriptsize
\caption{Simulation: Performance metrics for continuous-outcome scenarios at $q=100$, $r=10$, and $n_1=7000$. 
}
\label{append:tab:gaussian_q100_n17000}
\begin{tabular}{l|rrr|rrr}
\toprule
  & GLM & RRR & HiRRR & GLM & RRR & HiRRR \\
\midrule
  & \multicolumn{3}{c}{$b=0.01$}  & \multicolumn{3}{c}{$b=0.05$}  \\
$\mathrm{Er}(\bU_C)\times10^2$ & 12.80 (0.49) & 12.64 (0.48) & {\bf{12.47}} (0.47) & 0.53 (0.03) & 0.53 (0.03) & {\bf{0.53}} (0.03)\\
$\mathrm{Er}(\bV_C)\times10^2$ & 4.42 (0.25) & 3.68 (0.21) & {\bf{1.05}} (0.06) & 0.17 (0.01) & 0.14 (0.01) & {\bf{0.03}} (0.00)\\
$\mathrm{Er}(\bC)\times10^4$ & 5.89 (0.04) & 0.77 (0.01) & {\bf{0.64}} (0.01) & 5.89 (0.04) & 0.74 (0.01) & {\bf{0.62}} (0.01)\\
$\mathrm{Er}(\bbeta)\times10^4$ & 5.78 (0.38) & 1.39 (0.14) & {\bf{1.28}} (0.15) & 5.84 (0.31) & 1.35 (0.15) & {\bf{1.25}} (0.14)\\
$\mbox{Pred}(\bC)\times10^2$ & 15.01 (0.08) & 2.04 (0.03) & {\bf{1.65}} (0.03) & 15.00 (0.09) & 1.96 (0.03) & {\bf{1.60}} (0.03)\\
$\mbox{Pred}(\bbeta)\times10^2$ & 14.84 (0.93) & 3.61 (0.38) & {\bf{3.26}} (0.39) & 14.95 (0.77) & 3.52 (0.37) & {\bf{3.22}} (0.36)\\
\midrule
  & \multicolumn{3}{c}{$b=0.10$}  & \multicolumn{3}{c}{$b=0.15$}  \\
$\mathrm{Er}(\bU_C)\times10^2$ & 0.13 (0.01) & 0.13 (0.01) & {\bf{0.13}} (0.01) & 0.06 (0.00) & 0.06 (0.00) & {\bf{0.06}} (0.00)\\
$\mathrm{Er}(\bV_C)\times10^2$ & 0.04 (0.00) & 0.04 (0.00) & {\bf{0.01}} (0.00) & 0.02 (0.00) & 0.02 (0.00) & {\bf{0.00}} (0.00)\\
$\mathrm{Er}(\bC)\times10^4$ & 5.89 (0.04) & 0.74 (0.01) & {\bf{0.62}} (0.01) & 5.88 (0.04) & 0.74 (0.01) & {\bf{0.62}} (0.01)\\
$\mathrm{Er}(\bbeta)\times10^4$ & 5.83 (0.38) & 1.41 (0.16) & {\bf{1.31}} (0.17) & 6.02 (0.39) & 1.43 (0.16) & {\bf{1.32}} (0.16)\\
$\mbox{Pred}(\bC)\times10^2$ & 15.02 (0.08) & 1.95 (0.03) & {\bf{1.60}} (0.03) & 15.00 (0.09) & 1.95 (0.03) & {\bf{1.60}} (0.02)\\
$\mbox{Pred}(\bbeta)\times10^2$ & 14.84 (0.89) & 3.65 (0.41) & {\bf{3.34}} (0.42) & 15.33 (0.91) & 3.70 (0.42) & {\bf{3.38}} (0.40)\\
\bottomrule
\end{tabular}

\end{table}

\subsection{Simulation under General Encoder-Decoder Framework}\label{app:sim:nn}

We further investigated non-linear scenarios under the general encoder-decoder framework in Section \ref{sec:model:1}.

\begin{figure}[htp]
    \centering
    \begin{subfigure}[t]{0.45\textwidth}
        \centering
        \begin{tikzpicture}[node distance=32pt, decoration={calligraphic brace, mirror, amplitude=6pt}, font=\scriptsize]
    \pgfmathsetmacro{\nodedist}{32pt} 
    \pgfmathsetmacro{\distvectical}{0.6*\nodedist}
    
    \definecolor{darkpurple}{HTML}{9673A6}
    \definecolor{lightpurple}{HTML}{E1D5E7}
    \definecolor{darkblue}{HTML}{6C8EBF}
    \definecolor{lightblue}{HTML}{DAE8FC}
    \definecolor{darkgreen}{HTML}{82B366}
    \definecolor{lightgreen}{HTML}{D5E8D4}
    \definecolor{darkyellow}{HTML}{D6B656}
    \definecolor{lightyellow}{HTML}{FFF2CC}
    \definecolor{darkred}{HTML}{B85450}
    \definecolor{lightred}{HTML}{F8CECC}
    \definecolor{gray}{HTML}{666666}
    
    \tikzstyle{Node} = [circle, minimum height=22pt, align=center, draw=black, draw=gray, fill=gray!5, text=black]]
    \tikzstyle{dense} = [rectangle, rounded corners, minimum height=12pt, align=center, draw=black, draw=darkblue, fill=lightblue!50, text=black]]
    \tikzstyle{batchnorm} = [rectangle, rounded corners, minimum height=12pt, align=center, draw=black, draw=darkpurple, fill=lightred!50, text=black]]
    \tikzstyle{relu} = [rectangle, rounded corners, minimum height=12pt, align=center, draw=black, draw=darkyellow, fill=lightyellow!50, text=black]]
    \tikzstyle{sigmoid} = [rectangle, rounded corners, minimum height=12pt, align=center, draw=black, draw=darkyellow, fill=lightpurple!50, text=black]]
    \tikzstyle{dropout} = [rectangle, rounded corners, minimum height=12pt, align=center, draw=black, draw=darkgreen, fill=lightgreen!50, text=black]]
    \tikzstyle{outer} = [rectangle, rounded corners, draw=lightgray, thick, inner sep=4pt]
    \tikzstyle{arrow} = [thick,->,>=stealth]

    \node(dense1) [dense]{Dense Layer};
    \node(batchnorm1) [batchnorm, below of=dense1, node distance = \distvectical]{BatchNorm1d};
    \node(relu1) [relu, below of=batchnorm1, node distance = \distvectical]{ReLU};
    \node(dropout1) [dropout, below of=relu1, node distance = \distvectical]{Dropout};
    \node(outer1) [outer, fit=(dense1)(dropout1)] {};

    \node(dense2) [dense, right of = batchnorm1, node distance=2.4*\nodedist]{Dense Layer};
    \node(batchnorm2) [batchnorm, below of=dense2, node distance = \distvectical]{BatchNorm1d};
    \node(outer2) [outer, fit=(dense2)(batchnorm2)] {};
    \node[Node] (X) at ([xshift=-1.8*\nodedist]$(dense1)!.5!(dropout1)$) {in};
    \node[Node] (L) at ([xshift=1.8*\nodedist]$(dense2)!.5!(batchnorm2)$) {out};
    
    \draw[->] (dense1) -- (batchnorm1);
    \draw[->] (batchnorm1) -- (relu1);
    \draw[->] (relu1) -- (dropout1);
    \draw[->] (dense2) -- (batchnorm2);

    \draw[->] (X) -- (outer1.west);
    \draw[->] (outer1.east) -- (outer2.west);
    \draw[->] (outer2.east) -- (L);
    \draw[->] (outer1.south) -- ([yshift=-2.4*\distvectical]$(outer1.north)!.5!(outer1.south)$) -- ([yshift=-2.4*\distvectical]$(X.east)!.5!(outer1.west)$) -- ($(X.east)!.5!(outer1.west)$);
    \coordinate (text1) at ([yshift=-2.4*\distvectical]$(X.east)!.5!(outer1)$);
    \node[below] at (text1) {Repeat $k$ times.};
    \node[below, yshift=-0.5*\distvectical] at (X) {($p/q$)};
    \node[below, yshift=-0.5*\distvectical] at (L) {($r$)};
    \end{tikzpicture}
    \caption{Structure of the encoders}
    \end{subfigure}%
    \hspace{0.05\textwidth} %
    \begin{subfigure}[t]{0.45\textwidth}
        \centering
        \begin{tikzpicture}[node distance=32pt, decoration={calligraphic brace, mirror, amplitude=6pt}, font=\scriptsize]
    \pgfmathsetmacro{\nodedist}{32pt} 
    \pgfmathsetmacro{\distvectical}{0.6*\nodedist}
    
    \definecolor{darkpurple}{HTML}{9673A6}
    \definecolor{lightpurple}{HTML}{E1D5E7}
    \definecolor{darkblue}{HTML}{6C8EBF}
    \definecolor{lightblue}{HTML}{DAE8FC}
    \definecolor{darkgreen}{HTML}{82B366}
    \definecolor{lightgreen}{HTML}{D5E8D4}
    \definecolor{darkyellow}{HTML}{D6B656}
    \definecolor{lightyellow}{HTML}{FFF2CC}
    \definecolor{darkred}{HTML}{B85450}
    \definecolor{lightred}{HTML}{F8CECC}
    \definecolor{gray}{HTML}{666666}
    
    \tikzstyle{Node} = [circle, minimum height=22pt, align=center, draw=black, draw=gray, fill=gray!5, text=black]]
    \tikzstyle{dense} = [rectangle, rounded corners, minimum height=12pt, align=center, draw=black, draw=darkblue, fill=lightblue!50, text=black]]
    \tikzstyle{batchnorm} = [rectangle, rounded corners, minimum height=12pt, align=center, draw=black, draw=darkpurple, fill=lightred!50, text=black]]
    \tikzstyle{relu} = [rectangle, rounded corners, minimum height=12pt, align=center, draw=black, draw=darkyellow, fill=lightyellow!50, text=black]]
    \tikzstyle{sigmoid} = [rectangle, rounded corners, minimum height=12pt, align=center, draw=black, draw=darkyellow, fill=lightpurple!50, text=black]]
    \tikzstyle{dropout} = [rectangle, rounded corners, minimum height=12pt, align=center, draw=black, draw=darkgreen, fill=lightgreen!50, text=black]]
    \tikzstyle{outer} = [rectangle, rounded corners, draw=lightgray, thick, inner sep=4pt]
    \tikzstyle{arrow} = [thick,->,>=stealth]
    
    \node(dense3) at (0,3) [dense]{Dense Layer};
    \node(batchnorm3) [batchnorm, below of=dense3, node distance = \distvectical]{BatchNorm1d};
    \node(relu3) [relu, below of=batchnorm3, node distance = \distvectical]{ReLU};
    \node(dropout3) [dropout, below of=relu3, node distance = \distvectical]{Dropout};
    \node(outer3) [outer, fit=(dense3)(dropout3)] {};

    \node(dense4) [dense, right of = batchnorm3, node distance=2.4*\nodedist]{Dense Layer};
    \node(sigmoid4) [sigmoid, below of=dense4, node distance = \distvectical]{Sigmoid};
    \node(outer4) [outer, fit=(dense4)(sigmoid4)] {};
    \node[Node] (Y) at ([xshift=-1.8*\nodedist]$(dense3)!.5!(dropout3)$) {in};
    \node[Node] (Yr) at ([xshift=1.8*\nodedist]$(dense4)!.5!(sigmoid4)$) {out};
    
    \draw[->] (dense3) -- (batchnorm3);
    \draw[->] (batchnorm3) -- (relu3);
    \draw[->] (relu3) -- (dropout3);
    \draw[->] (dense4) -- (sigmoid4);

    \draw[->] (Y) -- (outer3.west);
    \draw[->] (outer3.east) -- (outer4.west);
    \draw[->] (outer4.east) -- (Yr);
    \draw[->] (outer3.south) -- ([yshift=-2.4*\distvectical]$(outer3.north)!.5!(outer3.south)$) -- ([yshift=-2.4*\distvectical]$(Y.east)!.5!(outer3.west)$) -- ($(Y.east)!.5!(outer3.west)$);
    
    \coordinate (text2) at ([yshift=-2.4*\distvectical]$(Y.east)!.5!(outer3)$);
    \node[below] at (text2) {Repeat $k$ times};
    \node[below, yshift=-0.5*\distvectical] at (Y) {($r$)};
    \node[below, yshift=-0.5*\distvectical] at (Yr) {($q$)};
    \end{tikzpicture}
    \caption{Structure of the decoders}
    \end{subfigure}
    \caption{Simulation: Structures of the neural networks for data generation. Note that $k$ is the number of compositional hidden layers, and the dimensions of the $k$ dense layers could vary; when $k=0$, it becomes a linear reduced-rank model. }
    \label{sim:fig:true_structure}
\end{figure}
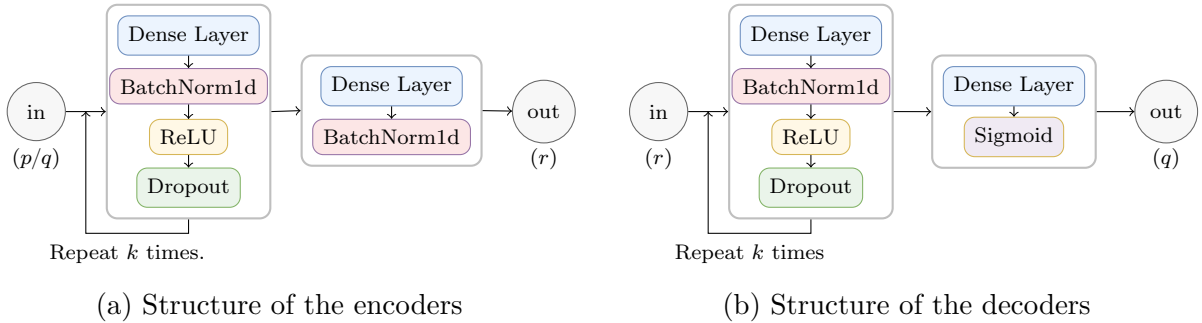

We constructed the true underlying models for $\mathcal H$, $\widetilde{\mathcal H}$, and $\mathcal G$ using neural networks, as illustrated in Figure~\ref{sim:fig:true_structure}. Each compositional layer was configured as a linear transformation followed by a BatchNorm1d layer (to maintain consistent distributions across intermediate layers), a ReLU activation function (to introduce non-linearity), and a dropout layer (to mitigate overfitting). The input $\bX$ was generated with entries drawn independently from $\mbox{N}(0,1)$, and the biases and weights of the neural networks were initially generated from $\mbox{N}(0,1)$. To align with the real application, we empirically adjusted the overall magnitude of the coefficients so that the expected AUC was around 70\%, and we further adjusted the final layer's bias term so that the prevalence of the primary outcome was roughly 20\%. The other settings were summarized as follows.

\begin{enumerate}\label{sim:binary_nonlinear_settings}
\item[\textbf{Scenario 3}] \label{binary_dl}
    (Binary outcomes with neural networks)    
    $(n,n_1)=(2000,7000)$ or $(20000,70000)$, $p=300$, $q=10$, $q_0=1$, $r=3$. The number of hidden units of $\mathcal H, \widetilde{\mathcal H}, \widetilde{\mathcal G} = [10,10], [3,3], [3,3]$, respectively.
\end{enumerate}

Several methods were considered, including GLM, RRR, HiRRR, and their neural network (NN) counterparts, namely, univariate neural networks for the primary outcome (Uni-NN), RRR with encoder-decoder networks (RRR-NN), and HiRRR with encoder-decoder networks (HiRRR-NN). For each NN-based method, we fixed the number of compositional layers at \(k=1\) to avoid overparameterization, while for the linear methods (GLM, RRR, and HiRRR) we set \(k=0\). We used a learning rate of 0.05, a batch size of 1000, and 200 training epochs, with a dropout rate of 0.2 in each encoder/decoder. A learning-rate scheduler reduced the learning rate by a factor of 0.6 every 50 epochs. For the smaller-sample scenario, we also applied $\ell_2$ regularization (weight decay $=10^{-4}$) to control overfitting.

Table \ref{sim:tab:binary_nn_metric} presents the key prediction metrics. Overall, the HiRRR methods consistently outperformed their RRR counterparts, although the degree of improvement varied by sample size. Notably, in the smaller dataset \((n,n_1)=(2000,7000)\), the linear HiRRR variant (i.e., without additional compositional layers) achieved the best performance, suggesting that a simpler model may suffice when sample size is more limited. In contrast, for the larger dataset \((n,n_1)=(20000,70000)\), allowing for nonlinear structures led to much stronger performance gains, indicating that neural network–based approaches can better capture complex relationships given sufficient data.

\begin{table}[htp]
\scriptsize
\centering
\caption{Simulation: Performance metrics for the general encoder-decoder scenarios. 
}
\label{sim:tab:binary_nn_metric}
\begin{tabular}[t]{lrrrr}
\toprule
\multirow{2}{*}{Model}&\multirow{2}{*}{AUC}&\multirow{2}{*}{PRAUC}&\multicolumn{2}{c}{90\% specificity} \\
&&& \makecell[c]{Sensitivity (\%)} & \makecell[c]{PPV (\%)} \\ 
\midrule
&\multicolumn{4}{c}{$(n,n_1)=(2000,7000)$}\\
GLM & 66.63 (5.92) & 34.01 (7.39) & 26.06 (8.46) & 38.38 (7.80) \\
Uni-NN & 63.28 (5.13) & 26.40 (4.84) & 21.53 (6.64) & 34.16 (6.92) \\
RRR & 69.16 (5.55) & 36.42 (7.77) & 28.68 (8.70) & 40.71 (7.36) \\
RRR-NN & 64.04 (4.66) & 29.52 (8.44) & 23.11 (7.77) & 35.58 (7.42) \\
HiRRR & {\bf{69.54}} (5.48) & {\bf{36.61}} (7.84) & {\bf{29.04}} (8.80) & {\bf{41.00}} (7.30) \\
HiRRR-NN & 64.32 (4.67) & 29.30 (8.26) & 23.34 (7.55) & 35.84 (7.50) \\
\midrule
&\multicolumn{4}{c}{$(n,n_1)=(20000,70000)$}\\
GLM & 73.60 (5.82) & 43.59 (10.97) & 36.59 (12.15) & 46.41 (8.50) \\
Uni-NN & 67.63 (6.33) & 13.87 (4.11) & 23.92 (10.29) & 35.88 (9.31) \\
RRR & 74.09 (5.54) & {\bf{43.99}} (10.80) & 37.03 (12.07) & 46.75 (8.34) \\
RRR-NN & 75.34 (6.40) & 42.98 (16.43) & 38.87 (15.60) & 47.19 (10.44) \\
HiRRR & 74.11 (5.62) & 43.68 (10.85) & 36.62 (12.18) & 46.43 (8.43) \\
HiRRR-NN & {\bf{75.36}} (5.89) & 43.70 (15.47) & {\bf{39.95}} (14.67) & {\bf{48.17}} (9.76) \\
\bottomrule
\end{tabular}

\end{table}

\FloatBarrier
\clearpage
\section{Proofs}

\subsection{Explicit Solution of HiRRR with Gaussian Outcomes}\label{sec: solution_lm}

In this section, we show the explicit solution of our HiRRR model with multivariate Gaussian outcomes. By optimizing the equation $\|\bY-\bX\bA\bB\trans\|_F^2 + \lambda\|\widetilde {\bY}-\widetilde{\bL}\bB\trans|_F^2 \text{ s.t. }\bB\trans\bB=\bI_r$ with respect to $\widetilde{\bL}$, the objective function becomes
\begin{align*}
& \|\bY-\bX\bA\bB\trans\|_F^2 + \lambda\|\widetilde {\bY}(\bI-\bB\bB\trans)|_F^2\\
 = & \|\bY\bB\bB\trans-\bX\bA\bB\trans\|_F^2 + \|\bY(\bI-\bB\bB\trans)\|_F^2 + \lambda\|\widetilde {\bY}(\bI-\bB\bB\trans)|_F^2\\
 = & \|\bY\bB-\bX\bA\|_F^2 + \|\bY(\bI-\bB\bB\trans)\|_F^2 + \lambda\|\widetilde {\bY}(\bI-\bB\bB\trans)|_F^2.
\end{align*}
Compared with the reduced rank regression with only $(\bX,\bY)$ considered, the additional term in the objective function is $\|\widetilde {\bY}(\bI-\bB\bB\trans)\|_F^2$,
which evaluates the projection of $\widetilde {\bY}$ onto the orthogonal complement of $\bB$, and equals to $\|\widetilde {\bE}(\bI-\bB\bB\trans)\|_F^2$ under the correct model assumption.

After optimizing the formal objective function with respect to $\A$, the objective problem becomes
\begin{align*}
\min_{B}\quad &\|\bP_X^\perp \bY\bB\|_F^2 + \|\bY(\bI-\bP_B)\|_F^2 + \lambda \|\widetilde {\bY}(\bI-\bP_B)\|_F^2\\
\Leftrightarrow \min_{B}\quad &\|\bP_X^\perp \bY\bB\|_F^2 - \|\bY\bB\|_F^2 - \lambda\|\widetilde {\bY}\bB\|_F^2\\
\Leftrightarrow \max_{B}\quad &\|\bP_X \bY\bB\|_F^2 + \|\widetilde {\bY}\bB\|_F^2.
\end{align*}
Denote the SVD result of the augmented matrix $\begin{pmatrix}\bP_X\bY\\\sqrt{\lambda}\widetilde {\bY}\end{pmatrix}=\bU\bD\bV\trans$. Then the optimal solution for $\bB$ is the first $r$ column vectors in $\bV$, denoted as $\bV_r$.

Then we could get the estimators of $\bA,\bB$ as
$$
\begin{cases}
\widehat {\bB}=\bV_r \\ 
\widehat {\bA}=(\bX\trans\bX)^{+}\bX\trans\bY\widehat {\bB}
\end{cases}
$$
The low rank matrix $\bC=\bA\bB\trans$ is estimated as $\widehat {\bC}_{HiR}=(\bX\trans \bX)^{+}\bX\trans \bY \bV_r\bV_r\trans$, 
while the reduced rank estimator based on complete dataset $(\bX,\bY)$ is $\widehat {\bC}_{R}=(\bX\trans \bX)^{+}\bX\trans \bY  \bV_r^o{\bV_r^o}\trans$, where ${\bV_r^o}$ is the first r right singular vectors of $\bP_X\bY$.

\subsection{Technique Assumptions \& Proof of Theorem~\ref{thm:HiRRR-Subspace}}\label{sec:proof}

We make the following assumptions. 
\begin{assumption}[Sub-Gaussian Noise]\label{as:1}  
Each row of ${\bE}$ is i.i.d.\ sub‐Gaussian with mean zero and positive definite covariance matrix $\bSigma_e$, each row of $\widetilde{{\bE}}$ is i.i.d.\ sub‐Gaussian with mean zero and positive definite covariance matrix $\widetilde{\bSigma}_{e}$, and ${\bE}$ and $\widetilde{{\bE}}$ are independent. 
\end{assumption}

\begin{assumption}[Multi-record Signal]\label{as:2}  
We assume $\|{\bA}\strans\,{\bx}_i\| = O(1)$, $i=1,\ldots,n$, where ${\bx}_i\trans\in \mathbb{R}^p$ is the $i$th row of ${\bX}$, so that ${\bA}\strans\,\bigl({\bX}\trans {\bX}\bigr)\,{\bA}^\star
\succeq c_1 n \bI_r$ for some $c_1>0$, and $\|{\bX}{\bA}^\star\|_{\mathrm{op}} = O(\sqrt{n})$.
\end{assumption}

\begin{assumption}[Single-record Signal]\label{as:3}    
We assume 
$\widetilde{{\bL}}^\star\in\mathbb{R}^{n_1\times r}$ has i.i.d.\ sub‐Gaussian rows $\widetilde{{\bl}}_j^\star$, $j=1,\ldots, n_1$, each with mean zero and covariance matrix $\bSigma_L$ satisfying $\bSigma_L\succeq c_1 \bI_r$, so that $\mathbb{E}[\widetilde{\bL}^\star\trans \widetilde{\bL}^\star] \succeq c_1 n_1 \bI_r$ and 
$\|\widetilde{{\bL}}^\star\|_{\mathrm{op}} =  O_p(\sqrt{n_1})$.  Also, $\widetilde{{\bL}}^\star$ is independent of the error matrices ${\bE}$ and $\widetilde{{\bE}}$. 
\end{assumption}

\begin{assumption}[Noise Separation] \label{as:4} 
Let $\bQ^\star = \bI_q - \bB^\star\bB\strans$. We assume 
$\bQ^\star 
(r_x\,\bSigma_e 
  \;+\;
  \lambda n_1\,\widetilde{\bSigma}_e)
\bQ^\star \preceq c_2 (n + \lambda n_1) \bI_{q-r}$, with $c_2 < c_1$, and $\|\bQ^\star 
(r_x\,\bSigma_e 
  \;+\;
  \lambda n_1\,\widetilde{\bSigma}_e)\bB^\star\bB\strans\|_{\mathrm{op}} = o(\sqrt{nq +\lambda n_1q})$.
\end{assumption}

The sub‐Gaussian noise assumptions \eqref{as:1} ensure that we can control operator‐norm fluctuations via matrix concentration, while \eqref{as:2} --\eqref{as:3} stipulate that both the multi‐record and single‐record data provide sufficiently strong signals in their respective rank‐$r$ structures.  Additionally, \eqref{as:4} enforces a ``noise separation,'' guaranteeing that $\mathbf{B}^\star$ remains the top‐$r$ eigenvector subspace of the population version of $\widehat{\bM}$. 


The following concentration inequalities for random vectors and matrices \citep{Tropp2012} are needed.

\begin{lemma}[Matrix Bernstein Inequality]
\label{lem:bernstein}
For a sequence $\{\bZ_k\}$ of $q_1\times q_2$ random satisfyning
$$
  \mathbb{E}[{\bZ_k}] 
  \;=\; 
  \0,
  \qquad 
  \bigl\|{\bZ_k}\bigr\|_{\mathrm{op}} \le R 
  \quad \textit{almost surely},
$$
the following inequality holds for all $t \ge 0$:
$$
  \Pr\!\Bigl(\bigl\|\sum_k {\bZ_k}\bigr\|_{\mathrm{op}} \ge t
  \Bigr)
  \;\le\;
  (q_1 + q_2)\cdot \exp \Bigl(\frac{-t^2/2}{\sigma^2 + Rt/3}
  \Bigr)
$$
where the variance parameter 
$$
  \sigma^2 
  \;=\;
  \max \bigl\{
  \bigl\| \sum_{k} \mathbb{E}({\bZ_k\bZ_k\trans}) \bigr\|_{\mathrm{op}},\; 
  \bigl\| \sum_{k} \mathbb{E}({\bZ_k\trans\bZ_k}) \bigr\|_{\mathrm{op}}
  \bigr\}.
$$
\end{lemma}

\begin{lemma}[Operator Norm of Sub-Gaussian Vector]
\label{lem:op_norm_subg}
Let \(\be \sim \mathrm{subG}(\0, \bSigma)\) be a sub-Gaussian random vector with mean \(\0\) and covariance matrix \(\bSigma \in \mathbb{R}^{p \times p}\). For any \(\bu \in \mathbb{R}^p\), the tail probability satisfies  
\[
\Pr\bigl(|\bu^\trans \be| \geq t\bigr) \leq \exp\bigl(-\frac{t^2}{2 \|\bSigma\|_{\mathrm{op}}^2 \|\bu\|_2^2}\bigr),
\]  
where \(\|\bSigma\|_{\mathrm{op}}\) denotes the operator norm of \(\bSigma\). Consequently, with high probability \(1 - 1/p\) or \(1 - 1/(np)\), the following bounds hold:
\[
\|\be\| \lesssim \|\bSigma\|_{\mathrm{op}} \log(p), \quad  
\max_{i=1,\dots,n} \|\be_i\| \lesssim \|\bSigma\|_{\mathrm{op}} \log(np),
\]  
where \(\be_1, \dots, \be_n\) are \(n\) independent samples from \(\mathrm{subG}(\0, \bSigma)\).
\end{lemma}


We first study the noise fluctuation of $\widehat{\bM}$. Recall that 
$$
  \widehat{\bM}
  \;=\;
  {\bY}\trans\,{\bP}_X\,{\bY}
  \;+\;
  \lambda \;\widetilde{{\bY}}\trans\,\widetilde{{\bY}},
  \qquad
  {\bP}_X
  \;=\;
  {\bX}\,({\bX}\trans {\bX})^+\,{\bX}\trans.
$$
From Proposition \ref{prop:1}, we know that the estimated loading matrix $\widehat{{\bB}}$ is the top‐$r$ eigenvectors of $\widehat{\bM}$.  Consider $\bM_{\mathrm{pop}}$, the \emph{population} counterpart of $\widehat{\bM}$, 
$$
  \bM_{\mathrm{pop}}
  \;=\;
  \mathbb{E}\bigl[{\bY}\trans\,{\bP}_X\,{\bY}\bigr]
  \;+\;
  \lambda\;\mathbb{E}\bigl[\widetilde{{\bY}}\trans\,\widetilde{{\bY}}\bigr].
$$

We will first bound $\|\widehat{\bM} - \bM_{\mathrm{pop}}\|_{\mathrm{op}}$. Consider 
$$
  \widehat{\bM} - \bM_{\mathrm{pop}}
  \;=\;
  \bigl({\bY}\trans\,{\bP}_X\,{\bY}
      \;-\;\mathbb{E}[{\bY}\trans\,{\bP}_X\,{\bY}]\bigr)
  \;+\;
  \lambda\,\bigl(\widetilde{{\bY}}\trans\,\widetilde{{\bY}}
                - \mathbb{E}[\widetilde{{\bY}}\trans\,\widetilde{{\bY}}]\bigr)
  \;=\;
  \Delta_1 \;+\; \lambda\,\Delta_2,              
$$
where
$$
  \Delta_1
  \;=\;
  {\bY}\trans\,{\bP}_X\,{\bY}
  \;-\;\mathbb{E}[{\bY}\trans\,{\bP}_X\,{\bY}],
  \qquad
  \Delta_2
  \;=\;
  \widetilde{{\bY}}\trans\,\widetilde{{\bY}}
  \;-\;\mathbb{E}[\widetilde{{\bY}}\trans\,\widetilde{{\bY}}].
$$

\noindent \textbf{Bounding $\|\Delta_1\|_{\mathrm{op}}$}:\\

First, since ${\bY} = {\bX}\,{\bA}^\star\,{\bB}\strans + {\bE}$, we have that 
$$
  {\bY}\trans {\bP}_X {\bY}
  \;=\;
  \bigl({\bB}^{\star}\,{\bA}\strans\,{\bX}\trans + {\bE}\trans\bigr)\,
  {\bP}_X \,
  \bigl({\bX}\,{\bA}^\star\,{\bB}\strans + {\bE}\bigr).
$$
Expanding this product yields four terms:
$$
  {\bY}\trans {\bP}_X {\bY}
  \;=\;
  \underbrace{{\bB}^{\star}\,{\bA}\strans\,{\bX}\trans {\bX}\,{\bA}^\star\,{\bB}\strans}
             _{(\mathrm{I})}
  \;+\;
  \underbrace{{\bB}^\star\,{\bA}\strans\,{\bX}\trans {\bE}}
             _{(\mathrm{II})}
  \;+\;
  \underbrace{{\bE}\trans {\bX}\,{\bA}^\star\,{\bB}\strans}
             _{(\mathrm{III})}
  \;+\;
  \underbrace{{\bE}\trans {\bP}_X {\bE}}
             _{(\mathrm{IV})}.
$$
Hence,
\begin{align*}
  \Delta_1 
  &\;=\;
  \bigl[(\mathrm{I}) + (\mathrm{II}) + (\mathrm{III}) + (\mathrm{IV})\bigr]
  \;-\;
  \mathbb{E}\bigl[(\mathrm{I}) + (\mathrm{II}) + (\mathrm{III}) + (\mathrm{IV})\bigr]\\
  &\;=\;
  \underbrace{ \Bigl[(\mathrm{I}) - \mathbb{E}(\mathrm{I})\Bigr] }_{=:\,\Delta_{1,\mathrm{I}}}
  \;+\;
  \underbrace{ \Bigl[(\mathrm{II}) - \mathbb{E}(\mathrm{II})\Bigr] }_{=:\,\Delta_{1,\mathrm{II}}}
  \;+\;
  \underbrace{ \Bigl[(\mathrm{III}) - \mathbb{E}(\mathrm{III})\Bigr] }_{=:\,\Delta_{1,\mathrm{III}}}
  \;+\;
  \underbrace{ \Bigl[(\mathrm{IV}) - \mathbb{E}(\mathrm{IV})\Bigr] }_{=:\,\Delta_{1,\mathrm{IV}}}.
\end{align*}
Below, we discuss each of the four terms.

The first term $\Delta_{1,\mathrm{I}}$ is zero if the design matrix is considered as fixed. 
To bound the second term $\Delta_{1,\mathrm{II}}$, we can view ${\bB}^\star\,{\bA}\strans\,{\bX}\trans {\bE}$ as a sum of (random) rank‐1 matrices from each row of ${\bE}$. Specifically, denote by ${\bx}_i\trans\in \mathbb{R}^p$ the $i$th row of ${\bX}$ and ${\be}_i\trans \in \mathbb{R}^q$ the $i$th row of ${\bE}$.  Then
$$
  {\bB}^\star{\bA}\strans{\bX}\trans\,{\bE}
  \;=\;
  \sum_{i=1}^n
    {\bB}^\star 
    \,\bigl({\bA}\strans\,{\bx}_i\bigr)\,
    {\be}_i\trans
  \;\;=\;\;
  \sum_{i=1}^n
    \bigl({\bB}^\star\,{\bA}\strans\,{\bx}_i\bigr)
    \,{\be}_i\trans.
$$
Each summand ${\bZ}_i = {\bB}^\star\,({\bA}\strans\,{\bx}_i)\,{\be}_i\trans$ is a rank‐1 matrix in $\mathbb{R}^{r\times q}$. Then $\mathbb{E}[{\bZ}_i] 
  = {\0}$ and
$$
  \max_i\;\|{\bZ}_i\|_{\mathrm{op}}
  \;\le\;
  \max_i\; \bigl\|{\bB}^\star\,{\bA}\strans\,{\bx}_i\bigr\|\cdot
  \max_i\; \|{\be}_i\|
  \;\le\;
  \|\bX\bA\|_{\mathrm{op}}\cdot
  \max_i\; \|{\be}_i\|
  \;\lesssim\;
  \sqrt{n\,q}
$$
with large probability from Lemma~\ref{lem:op_norm_subg}.
Moreover, 
\begin{align*}
  \bigl\|\mathbb{E}\bigl[\sum_i \bZ_i\bZ_i\trans
  \bigr]\bigr\|_{\mathrm{op}}
  \;&=\;
  \bigl\|\mathbb{E}\bigl[\sum_i\bB^\star\bA\strans \bx_i \be_i\trans\be_i \bx_i^\trans\bA^\star\bB\strans
  \bigr]\bigr\|_{\mathrm{op}} \\
  \;&=\;
  \mathrm{Tr}(\bSigma_e)\bigl\|\bB^\star\bA\strans \mathbb{E}\bigl[\bX\trans\bX\bigr]\bA^\star\bB\strans
  \bigr\|_{\mathrm{op}} \\
  \;&\lesssim\;
  n\mathrm{Tr}(\bSigma_e)
  \;\lesssim\;
  nq,
\end{align*}
\begin{align*}
  \bigl\|\mathbb{E}\bigl[\sum_i \bZ_i\trans\bZ_i
  \bigr]\bigr\|_{\mathrm{op}}
  \;&=\;
  \bigl\|\mathbb{E}\bigl[\sum_i \be_i\bx_i^\trans\bA^\star\bA^\strans\bx_i\be_i\trans
  \bigr]\bigr\|_{\mathrm{op}} \\
  \;&=\;
  \mathbb{E}\bigl[\|\bX\bA^\star\|_F^2\bigr]\cdot
  \bigl\|\bSigma_e
  \bigr\|_{\mathrm{op}} \\
  \;&\lesssim\;
  n\|\bSigma_e\|_{\mathrm{op}}
  \;\lesssim\;
  n.
\end{align*}

With Lemma~\ref{lem:bernstein}, we can get
$$
  \Pr\!\Bigl(
    \bigl\|{\bB}^\star\,{\bA}\strans\,{\bX}\trans\,{\bE}\bigr\|_{\mathrm{op}}
    \;\ge\; 
    t
  \Bigr)
  \;\le \;
  2q \cdot \exp\Bigl(-\;\frac{t^2/2}{nq + \sqrt{nq}\,t/3}\Bigr),
$$
and
$$
  \Pr\!\Bigl(
    \bigl\|{\bB}^\star\,{\bA}\strans\,{\bX}\trans\,{\bE}\bigr\|_{\mathrm{op}}
    \;\lesssim\; 
    \sqrt{n\,q}\, {\color{blue} \log(n\,q)}
  \Bigr)
  \;> \;
  1-\frac{c_3}{n\,q},
$$
for some constant $c_3>0$. This implies that 
\begin{align}
  \|\Delta_{1,\mathrm{II}}\|_{\mathrm{op}}
  \;=\;
  O_{\mathbb{P}}\Bigl(\sqrt{n\,q}\Bigr),\label{eq:delta12}
 \end{align} 
ignoring the smaller log term. Analogously, the third term has exactly the same rate. 


For the fourth term, we can similarly write it as a sum of matrices
$$
   {\bE}\trans {\bP}_X {\bE} - \mathbb{E}\bigl[{\bE}\trans {\bP}_X {\bE}\bigr] 
   \;=\;
   \sum_i ({\bP}_X)_{i,i} \, ( {\be}_i{\be}_i\trans -\bSigma_e ).
$$
The operator norm of each summand
$$
  \max_i\; \bigl\| {(\bP}_X)_{i,i} \, ({\be}_i{\be}_i\trans -\bSigma_e) \bigr\|_{\mathrm{op}}
  \;\le\;
  \max_i\; \|{\be}_i\|^2 + \|\bSigma_e\|_{\mathrm{op}}
  \;\lesssim\;
  \log(nq),
$$
with probability at least $1-1/(np)$,
and the total variance
$$
  \bigl\|\mathbb{E}\bigl[\sum_i ({\bP}_X)_{i,i}^2 \; 
  ({\be}_i{\be}_i\trans -\bSigma_e)^2
  \bigr]\bigr\|_{\mathrm{op}}
  \;=\;
  \bigl\|\mathbb{E}\bigl[\sum_i ({\bP}_X)_{i,i}^2\bigr] \; 
  (\bSigma_e^2 - \mathrm{Tr}(\bSigma_e)\bSigma_e) 
  \bigr\|_{\mathrm{op}}
  \;\lesssim \;
  r_x\,\mathrm{Tr}(\bSigma_e)\, \|\bSigma_e\|_{\mathrm{op}},
  \;\lesssim \;
  r_x \,q.
$$
With Lemma~\ref{lem:bernstein}, we get 
$$
  \Pr\!\Bigl(
    \bigl\|{\bE}\trans {\bP}_X\,{\bE} - r_x\,\bSigma_e\|_{\mathrm{op}}
    \;\lesssim\; 
    \sqrt{n\,q}\log(n\,q)
  \Bigr)
  \;> \;
  1-\frac{c_4}{n\,q},
$$
for some constant $c_4$. Therefore,
\begin{align}
  \| \Delta_{1,\mathrm{IV}} \|_{\mathrm{op}}
  \;=\;
  O_{\mathbb{P}}\!\Bigl(\sqrt{n\,q}\Bigr).\label{eq:delta14}
\end{align}
Combining the results from \eqref{eq:delta12} and \eqref{eq:delta14}, we get that 
$$
  \|\Delta_1\|_{\mathrm{op}}
  \;=\;
  \|{\bY}\trans {\bP}_X {\bY} - \mathbb{E}[{\bY}\trans {\bP}_X {\bY}]\|_{\mathrm{op}}
  \;\;=\;\;
  O_{\mathbb{P}}\!\Bigl(\sqrt {n\,q}
  \Bigr).
$$

\noindent \textbf{Bounding $\|\Delta_2\|_{\mathrm{op}}$}:\\
The treatment of $\|\Delta_2\|_{\mathrm{op}}$ is similar to that of $\|\Delta_1\|_{\mathrm{op}}$. Based on the single-record model, we write 
$$
  \widetilde{{\bY}}\trans \widetilde{{\bY}}
  \;=\;
  \underbrace{{\bB}^\star\,\widetilde{{\bL}}\strans\,\widetilde{{\bL}}^\star\,({\bB}^\star)\trans}_{(\mathrm{I})}
  \;+\;
  \underbrace{{\bB}^\star\,\widetilde{{\bL}}\strans\,\widetilde{{\bE}}}_{(\mathrm{II})}
  \;+\;
  \underbrace{\widetilde{{\bE}}\trans\,\widetilde{{\bL}}^\star\,({\bB}^\star)\trans}_{(\mathrm{III})}
  \;+\;
  \underbrace{\widetilde{{\bE}}\trans \widetilde{{\bE}}}_{(\mathrm{IV})}.
$$
Simularly, we can rewrite $\Delta_2$ as a sum of four terms:
$$
  \Delta_2
  \;=\;
  \underbrace{ \Bigl[(\mathrm{I}) - \mathbb{E}(\mathrm{I})\Bigr] }_{=:\,\Delta_{2,\mathrm{I}}}
  \;+\;
  \underbrace{ \Bigl[(\mathrm{II}) - \mathbb{E}(\mathrm{II})\Bigr] }_{=:\,\Delta_{2,\mathrm{II}}}
  \;+\;
  \underbrace{ \Bigl[(\mathrm{III}) - \mathbb{E}(\mathrm{III})\Bigr] }_{=:\,\Delta_{2,\mathrm{III}}}
  \;+\;
  \underbrace{ \Bigl[(\mathrm{IV}) - \mathbb{E}(\mathrm{IV})\Bigr] }_{=:\,\Delta_{2,\mathrm{IV}}}.
$$

Consider the first term. The rows of $\widetilde{{\bL}}^\star$, $\widetilde{{\bl}}_j^\star \in \mathbb{R}^r$,  
are i.i.d. with 
$\mathbb{E}[\widetilde{{\bl}}_j^\star(\widetilde{{\bl}}_j^\star)\trans] = \bSigma_L$.  
Then 
$
\mathbb{E}\bigl[\widetilde{{\bL}}\strans\,\widetilde{{\bL}}^\star\bigr]
  \;=\;
  n_1 \,\bSigma_L
$.
We have that 
$$
  (\mathrm{I}) - \mathbb{E}(\mathrm{I})
  \;=\;
  {\bB}^\star\,
  \bigl(\widetilde{{\bL}}\strans\,\widetilde{{\bL}}^\star - n_1 \bSigma_L\bigr)\,
  {\bB}^\strans.
$$
Since with probability at least $1-1/(n_1r)$ ,
$$
  \max_{j} \|\widetilde{{\bl}}_j^\star(\widetilde{{\bl}}_j^\star)\trans -\bSigma_L\|_{\mathrm{op}}
  \;\le\;
  \max_{j} \|\widetilde{{\bl}}_j\|^2 + \|\bSigma_L\|_{\mathrm{op}}
  \;\lesssim\;
  \log^2(n_1 r),
$$
and
$$
  \bigl\|\bigr(\mathbb{E}\bigl[\widetilde{{\bL}}\strans\,\widetilde{{\bL}}^\star\bigl]-n_1\bSigma_L\bigr)^2
  \bigr\|_{\mathrm{op}}
  \;\lesssim\;
  n_1\, \mathrm{Tr}(\Sigma_L)\, \|\Sigma_L\|_{\mathrm{op}}
  \;\lesssim\;
  n_1\,r,
$$
with Lemma~\ref{lem:bernstein}, we can get
$$
   \Pr\!\Bigl(
    \bigl\|
    {\bB}^\star\,
    \bigl(\widetilde{{\bL}}\strans\,\widetilde{{\bL}}^\star - n_1 \bSigma_L\bigr)\,
    ({\bB}^\star)\trans
    \|_{\mathrm{op}}
    \;\lesssim\; 
    \sqrt{n_1\,r} \log(n_1\,r)
  \Bigr)
  \;> \;
  1-\frac{c_5}{n_1\,r},
$$
for some constant $c_5$. Hence,
\begin{align}
  \|\Delta_{2,\mathrm{I}}\|_{\mathrm{op}}
  \;=\;
  O_{\mathbb{P}}\!\Bigl(\sqrt{n_1\,r}\Bigr).\label{eq:delta21}
\end{align}

For the second term, we know that 
$
\mathbb{E}(\mathrm{II})
  \;=\;
\mathbb{E}
\bigl({\bB}^\star\,\widetilde{{\bL}}\strans\,\widetilde{{\bE}}\bigr) 
  \;=\; {\0}
$.
We can again write $(\mathrm{II})$ as a sum of random rank‐1 matrices and apply Lemma~\ref{lem:op_norm_subg} to get
$$
   \max_j \|\bB^\star\widetilde{{\bl}}_j^\star(\widetilde{{\be}}_j^\star)\trans\|_{\mathrm{op}}
  \;\lesssim\;
  \sqrt{n\cdot\max\{q, r\}}
$$
with probability at least $1-1/(n\cdot\max\{q, r\})$, and
\begin{align*}
  \bigl\|\mathbb{E}\bigl[\sum_j \widetilde{{\bl_j}}^\star\,\widetilde{{\be_j}}^\trans\,\widetilde{{\be_j}}\,\widetilde{{\bl_j}}\strans
  \bigl]\bigr\|_{\mathrm{op}}
  \;&\lesssim\;
  n_1 \mathrm{Tr}(\bSigma_e) \|\bSigma_L\|_{\mathrm{op}}
  \;\lesssim\;
  n_1\,q,\\
  \bigl\|\mathbb{E}\bigl[\sum_j 
  \widetilde{{\be_j}}\,\widetilde{{\bl_j}}\strans\,\,\widetilde{{\bl_j}}^\star\,\widetilde{{\be_j}}^\trans
  \bigl]\bigr\|_{\mathrm{op}}
  \;&\lesssim\;
  n_1 \mathrm{Tr}(\bSigma_L) \|\bSigma_e\|_{\mathrm{op}}
  \;\lesssim\;
  n_1\,r.
\end{align*}
Therefore,
\begin{align}
  \|\Delta_{2,\mathrm{II}}\|_{\mathrm{op}}
  \;=\;
  O_{\mathbb{P}}\!\Bigl(\sqrt{n_1\cdot\max\{q, r\}}\Bigr)
  \;=\;
  O_{\mathbb{P}}\!\Bigl(\sqrt{n_1\,q}\Bigr).\label{eq:delta22}
\end{align}
Also, the third term is of the same rate. 

Lastly, for the fourth term, 
\begin{align}
  \|\Delta_{2,\mathrm{IV}}\|_{\mathrm{op}}
  \;\lesssim\;
  O_{\mathbb{P}}\!\Bigl(\sqrt{n_1\,q}\Bigr),\label{eq:delta24}
\end{align}
since $\max_j \|\widetilde{{\be_j}}\,\widetilde{{\be_j}}^\strans - \bSigma_e\|_{\mathrm{op}} = O_{\mathbb{P}}(\sqrt{n_1\,q})$ and $\bigl\|\mathbb{E}\bigl[
  (\widetilde{{\bE}}\trans \widetilde{{\bE}} - n_1\widetilde{\bSigma}_e)^2
  \bigl]\bigr\|_{\mathrm{op}}\lesssim n_1 q$.

Combining results from \eqref{eq:delta21}, \eqref{eq:delta22}, and \eqref{eq:delta24}, we have 
$$
  \|\Delta_2\|_{\mathrm{op}}
  \;\le\;
  \|\Delta_{2,\mathrm{I}}\|_{\mathrm{op}}
  \;+\;
  \|\Delta_{2,\mathrm{II}}\|_{\mathrm{op}}
  \;+\;
  \|\Delta_{2,\mathrm{III}}\|_{\mathrm{op}}
  \;+\;
  \|\Delta_{2,\mathrm{IV}}\|_{\mathrm{op}}
  \;\;=\;\;
  O_{\mathbb{P}}\!\Bigl(\sqrt{n_1\, q}\Bigr).
$$
For simplicity, we have omitted the smaller logarithmic factors.\\ 

\noindent \textbf{Combining $\Delta_1$ and $\Delta_2$}:\\
By the triangle inequality,
$$
  \|\widehat{\bM} - \bM_{\mathrm{pop}}\|_{\mathrm{op}}
  \;\;\le\;\;
  \|\Delta_1\|_{\mathrm{op}}
  \;+\;\lambda\,\|\Delta_2\|_{\mathrm{op}}.
$$
Hence, a union bound implies that with probability at least 
$$
  1 - \frac{c_3}{n\,q} - \frac{c_4}{n_1\,q} - \frac{c_5}{n_1\,r},
$$
we have that,
%
$$
  \|\widehat{\bM} - \bM_\mathrm{pop}\|_{\mathrm{op}}
  \;=\;
  O_{\mathbb{P}}\!\Bigl(
    \sqrt {n\,q} + \lambda\,\sqrt{n_1\, q}
  \Bigr).
$$

\noindent \textbf{Subspace Estimation Error}:\\
We are interested in bounding the \emph{subspace estimation error}:
$$
  \bigl\|\sin\!\bigl(\Theta(\widehat{{\bB}}, {\bB}^\star)\bigr)\bigr\|_{\mathrm{op}}
  \;=\;
  \bigl\|\widehat{{\bB}}\widehat{{\bB}}\trans - {\bB}^\star ({\bB}^\star)\trans\bigr\|_{\mathrm{op}}.
$$

Recall that 
\begin{align*}
  \bM_{\mathrm{pop}}
  \;&=\;
  \mathbb{E}\bigl[{\bY}\trans\,{\bP}_X\,{\bY}\bigr] +
  \mathbb{E}[\widetilde{{\bY}}\trans \widetilde{{\bY}}]\\
  \;&=\;
  {\bB}^\star \bigl({\bA}\strans\,{\bX}\trans {\bX}\,{\bA}^\star\bigr)
  {\bB}\strans + 
  {\bB}^\star 
  \bigl(\mathbb{E}[\widetilde{{\bL}}\strans\,\widetilde{{\bL}}^\star]\bigr)\,
  ({\bB}^\star)\trans
  + r_x \bSigma_e
  + \lambda n_1 \widetilde{\bSigma}_e.
\end{align*}
Define 
\begin{align*}
  \bM
  \;&=\;
  {\bB}^\star \bigl({\bA}\strans\,{\bX}\trans {\bX}\,{\bA}^\star\bigr)
  {\bB}\strans + 
  {\bB}^\star 
  \bigl(\mathbb{E}[\widetilde{{\bL}}\strans\,\widetilde{{\bL}}^\star]\bigr)\,
  ({\bB}^\star)\trans\\
  &\quad +\bB^\star\bB\strans (r_x \bSigma_e + n_1 \widetilde{\bSigma}_e) \bB^\star\bB\strans 
  + \bQ^\star (r_x \bSigma_e + \lambda n_1 \widetilde{\bSigma}_e) \bQ^\star.
\end{align*}
Then, we could write $\bM_{\mathrm{pop}}$ as 
\begin{align*}
  \bM_{\mathrm{pop}}
  \;&=\;
  \bM +  \bB^\star\bB (r_x \bSigma_e + \lambda n_1 \widetilde{\bSigma}_e) \bQ^\star 
  + \bQ^\star(r_x \bSigma_e + \lambda n_1 \widetilde{\bSigma}_e)\bB^\star\bB.
\end{align*}

Assumptions \eqref{as:3}--\eqref{as:4} ensure that the true loading matrix ${\bB}^\star$ is the top‐$r$ eigenvectors of $\bM$, and moreover, the eigen-gap satisfies 
$\delta = \lambda_r(\bM) - \lambda_{r+1} (\bM) = O(n + \lambda\,n_1)$. Assumption \eqref{as:4} also ensures that $\|\bM-\bM_{\mathrm{pop}}\|_{\mathrm{op}} \le 2\|\bB^\star\bB\strans (r_x \bSigma_e + \lambda n_1 \widetilde{\bSigma}_e) \bQ^\star\|_{\mathrm{op}} = o(\sqrt{n\,q + \lambda n_1\,q})$. Therefore, based on the Davis--Kahan theorem \citep{YuSamwrth2014} and the triangle inequality, we have that 
\begin{align*}
  \|\sin\!\bigl(\Theta(\widehat{{\bB}}, {\bB}^\star)\bigr)\|_{\mathrm{op}}
  &\;\;\le\;\;
  \frac{2\,\|\widehat{\bM} - \bM\|_{\mathrm{op}}}{\delta}\\
  &\;\;\le\;\;
  \frac{
  2\,
  \| \widehat{\bM} - \bM_{\mathrm{pop}}
  \|_{\mathrm{op}} 
  + 2\, 
  \| {\bM_{\mathrm{pop}}} - \bM 
  \|_{\mathrm{op}}
  }{\delta}\\
  &\;\;\lesssim\;\;
     \frac{\sqrt{n\,q}
           + \lambda\,\sqrt{n_1\,q}}{\,n + \lambda\,n_1\,} \\       
  &\;\;\asymp\;\;
  \frac{\sqrt{q}}{\sqrt{\,n + \lambda\,n_1}},
\end{align*}
with probability at least
$$
  1 - \frac{c_3}{n\,q} - \frac{c_4}{n_1\,q} - \frac{c_5}{n_1\,r}.
$$

This completes the proof.


\clearpage
\section{Additional Results from Suicide Risk Study}\label{app:suicide}

\begin{table}[htb]
\centering
\tiny
\caption{Application: Top 10 protective predictors from fitting GLM, RRR, and HiRRR over 10 random splits. Also reported are the prevalence of each condition/exposure and the log odds ratio between the exposure and the outcome of suicide attempt.}
\label{tab:app:protective}
\begin{tabular}{l L{3.6cm} r r r r r}
\toprule
\makecell[l]{Variable} & \makecell[l]{Description} & \makecell[c]{Prevalence in\\case/control} & \makecell[c]{Log\\odds\\ratio} & \makecell[c]{HiRRR} & \makecell[c]{RRR} & \makecell[c]{GLM}\\
\midrule
ICD-9 754 & Certain congenital musculoskeletal deformities & 0.0\% / 0.8\% & -Inf & {\bf{-2.230}} (0.165) & -0.018 (0.003) & -0.837 (0.296)\\
ICD-9 V44 & Artificial opening status & 0.0\% / 2.0\% & -Inf & {\bf{-1.003}} (0.270) & -0.070 (0.016) & {\bf{-1.365}} (0.090)\\
ICD-9 753 & Congenital anomalies of urinary system & 0.0\% / 0.7\% & -Inf & {\bf{-0.869}} (0.060) & -0.015 (0.005) & -0.274 (0.441)\\
ICD-9 755 & Other congenital anomalies of limbs & 0.0\% / 0.9\% & -Inf & {\bf{-0.579}} (0.088) & -0.035 (0.008) & -0.916 (0.325)\\
ICD-9 V43 & Organ or tissue replaced by other means & 0.0\% / 0.7\% & -Inf & {\bf{-0.513}} (0.426) & -0.003 (0.003) & 0.000 (0.000)\\
ICD-9 191 & Malignant neoplasm of brain & 0.0\% / 0.7\% & -Inf & {\bf{-0.386}} (0.035) & -0.034 (0.011) & -0.090 (0.286)\\
ICD-9 742 & Other congenital anomalies of nervous system & 0.0\% / 1.1\% & -Inf & {\bf{-0.350}} (0.059) & -0.051 (0.014) & {\bf{-1.157}} (0.040)\\
ICD-9 759 & Other and unspecified congenital anomalies & 0.0\% / 0.7\% & -Inf & {\bf{-0.306}} (0.248) & -0.016 (0.007) & -0.090 (0.284)\\
ICD-9 585 & Chronic kidney disease (ckd) & 0.0\% / 0.7\% & -Inf & {\bf{-0.297}} (0.049) & -0.027 (0.005) & -0.384 (0.497)\\
ICD-9 273 & Disorders of plasma protein metabolism & 0.0\% / 0.7\% & -Inf & {\bf{-0.260}} (0.113) & -0.052 (0.015) & -0.282 (0.454)\\
ICD-9 540 & Acute appendicitis & 0.2\% / 2.8\% & -2.464 & -0.172 (0.020) & {\bf {-0.133}} (0.013) & -0.455 (0.533)\\
ICD-9 555 & Regional enteritis & 0.4\% / 3.0\% & -1.868 & -0.094 (0.030) & {\bf {-0.087}} (0.020) & -0.206 (0.055)\\
ICD-9 560 & Intestinal obstruction without mention of hernia & 0.8\% / 2.9\% & -1.157 & -0.142 (0.026) & {\bf {-0.086}} (0.016) & -0.013 (0.046)\\
ICD-9 299 & Pervasive developmental disorders & 4.7\% / 3.2\% & 0.315 & -0.095 (0.010) & {\bf {-0.084}} (0.009) & 0.003 (0.006)\\
ICD-9 493 & Asthma & 11.9\% / 15.4\% & -0.249 & -0.108 (0.015) & {\bf {-0.083}} (0.011) & -0.139 (0.077)\\
ICD-9 998 & Other complications of procedures not elsewhere classified & 0.6\% / 2.3\% & -1.184 & -0.098 (0.032) & {\bf {-0.082}} (0.017) & 0.068 (0.061)\\
Schizophrenia &  & 3.9\% / 2.8\% & 0.280 & -0.110 (0.044) & {\bf {-0.077}} (0.024) & -0.044 (0.029)\\
ICD-9 338 & Pain, not elsewhere classified & 2.3\% / 4.1\% & -0.545 & -0.139 (0.013) & {\bf {-0.076}} (0.010) & -0.039 (0.025)\\
ICD-9 590 & Infections of kidney & 0.2\% / 1.9\% & -2.073 & -0.128 (0.019) & {\bf {-0.073}} (0.009) & -0.325 (0.461)\\
ICD-9 782 & Symptoms involving skin and other integumentary tissue & 0.8\% / 3.3\% & -1.274 & -0.139 (0.013) & {\bf {-0.072}} (0.010) & -0.097 (0.021)\\
ICD-9 996 & Complications peculiar to certain specified procedures & 0.0\% / 2.7\% & -Inf & -0.171 (0.022) & -0.069 (0.020) & {\bf{-1.629}} (0.090)\\
ICD-9 512 & Pneumothorax and air leak & 0.0\% / 1.8\% & -Inf & -0.114 (0.011) & -0.048 (0.012) & {\bf{-1.528}} (0.063)\\
ICD-9 511 & Pleurisy & 0.0\% / 1.7\% & -Inf & -0.122 (0.041) & -0.036 (0.012) & {\bf{-1.317}} (0.079)\\
ICD-9 528 & Diseases of the oral soft tissues excluding lesions specific for gingiva and tongue & 0.0\% / 1.8\% & -Inf & 0.151 (0.031) & 0.042 (0.016) & {\bf{-1.236}} (0.068)\\
ICD-9 349 & Other and unspecified disorders of the nervous system & 0.0\% / 1.1\% & -Inf & 0.045 (0.011) & 0.008 (0.006) & {\bf{-1.228}} (0.043)\\
ICD-9 733 & Other disorders of bone and cartilage & 0.0\% / 1.3\% & -Inf & -0.081 (0.017) & -0.021 (0.007) & {\bf{-1.196}} (0.042)\\
ICD-9 478 & Other diseases of upper respiratory tract & 0.0\% / 1.1\% & -Inf & -0.136 (0.015) & -0.036 (0.011) & {\bf{-1.187}} (0.044)\\
ICD-9 578 & Gastrointestinal hemorrhage & 0.0\% / 1.1\% & -Inf & 0.016 (0.020) & -0.014 (0.003) & {\bf{-1.186}} (0.076)\\
\bottomrule
\end{tabular}

\end{table}

\end{document}